\documentclass[12pt,a4paper, fleqn]{article}

\RequirePackage[l2tabu, orthodox]{nag}

\usepackage{mjheppub}

\usepackage[british]{babel}
\usepackage[utf8x]{inputenc}
\usepackage{fancyhdr}
\usepackage{amsmath}
\usepackage{amssymb}
\usepackage{yfonts}
\usepackage[colorinlistoftodos]{todonotes}
\usepackage{color}
\usepackage{mathrsfs}
\usepackage{subcaption}
\DeclareCaptionLabelFormat{andtable}{#1~#2  \&  \tablename~\thetable}
\usepackage{braket}
\usepackage{xfrac}
\usepackage{dsfont}
\usepackage{amsthm,amssymb,amsmath,epic,eepic,float}
\usepackage{rotating,epsfig,indentfirst,array,varioref}
\usepackage{appendix,marginnote,bbm,tikz,pgf,mathtools}

\usepackage{lipsum}
\usepackage{amsthm,amssymb,amsmath,epic,eepic,float}
\usepackage{rotating,epsfig,indentfirst,array,varioref}
\usepackage{appendix,marginnote,bbm,tikz,pgf,mathtools}

\usepackage{tikz}
\usetikzlibrary{calc,tikzmark,shapes,decorations.pathmorphing,decorations.pathreplacing,decorations.text,decorations,shapes.symbols,arrows,positioning,backgrounds}
\usetikzlibrary{calc}
\usepackage{pgfplots}

\usepackage{empheq} 

\def\leq{\leqslant}
\def\geq{\geqslant}

\def\x{x^{(1)}}

\newcommand{\bea}{\begin{eqnarray}\displaystyle}
\newcommand{\eea}{\end{eqnarray}}

\catcode`@=12
\relax

\def\beq{\begin{equation}}
\def\eeq{\end{equation}}
\def\bea{\begin{eqnarray}}
\def\eea{\end{eqnarray}}
\def\EQ{\begin{equation}}
\def\EN{\end{equation}}

\relax
\usepackage{mathtools}

\numberwithin{figure}{section}
\numberwithin{table}{section}
\usepackage{etoolbox}
\AtBeginEnvironment{figure}{\stepcounter{table}}
\AtBeginEnvironment{table}{\stepcounter{figure}}
\title{\boldmath  Three- and four-point connectivities of two-dimensional critical $Q-$ Potts random clusters on the torus}
\author{Nina Javerzat$^1$, Marco Picco$^2$, Raoul Santachiara$^1$}
\affiliation{$^1$ LPTMS, CNRS (UMR 8626), Univ.Paris-Sud, Universit\'e Paris-Saclay, 91405 Orsay, France}
\affiliation{\vspace{2mm}
$^2$ LPTHE, UMR 7589, Sorbonne Universit\'e and CNRS, France}
\emailAdd{nina.javerzat@u-psud.fr, picco@lpthe.jussieu.fr, raoul.santachiara@u-psud.fr}
\abstract{
In a recent paper, we considered the effects of the torus lattice topology on the two-point connectivity of $Q-$ Potts clusters. These effects are universal and probe non-trivial structure constants of the theory. We complete here this work by considering the torus corrections to the three- and four-point connectivities. These corrections, which depend on the scale invariant ratios of the triangle and quadrilateral formed by the three and four given points, test other non-trivial structure constants. We also present results of Monte Carlo simulations in good
agreement with our predictions.
}
\preprint{\today}
\begin{document}
\maketitle
\flushbottom
\section{Introduction}
The two-dimensional $Q-$ Potts model is a one-parameter family of models which describe random clusters on a lattice \cite{Wupotts} and admit for $Q\leq 4$ a continuous transition between a percolating and a non-percolating cluster phase \cite{Duminil_pc10}. At the critical point the clusters form conformal invariant fractal structures whose description challenges our understanding of the fractal geometry in critical phenomena \cite{du89}. For more than thirty years, physicists have been trying to solve the Conformal Field Theories (CFT) that capture, for general $Q\in \mathbb{R}, Q\leq 4$, the continuum limit of the critical $Q-$ Potts models. Despite many important results, in particular the computation of the partition function \cite{fsz87}, the derivation of exact formulas for many critical exponents \cite{saleur87} and the progress in the representation theory of the Temperley-Lieb type algebras underlying these models \cite{ReSa, Gainutdinov_2013}, the problem of defining the correct CFT solution remains an open issue. In particular, the knowledge of the CFT structure constants, which determine the small-distance asymptotic behaviour of the CFT many-point correlation functions, is missing. 
\noindent A remarkable proposal was done in \cite{devi11}, where the so-called $c\leq 1$ Liouville  structure constants \cite{zam05,kp05a,sch05} $-$ until then considered non-physical $-$ were conjectured to describe the three-point connectivity, i.e. the probability that three given points belong to the same cluster. Inspired by this result, new crossing-symmetric solutions, based on the $c\leq 1$ Liouville-type constants, have been found \cite{RiSa14, prs16,mr17, prs19,ri19}, and some of them proposed to describe four-point cluster connectivities \cite{prs16,prs19}. In \cite{js18} it was argued, and numerically shown, that there are states which provide a non-vanishing contribution to the connectivities but that are not taken into account by these bootstrap solutions. These contributions are, for general values of $Q$, very small and the bootstrap solutions remain a very good approximation (within the Monte Carlo simulation precision) to the cluster connectivities.

\noindent In \cite{jps19two}, we considered the effects of the torus lattice topology on the two-point Potts connectivity, which probe non-trivial structure constants of the theory. Putting together the exact analysis in \cite{js18} and the results in \cite{prs16,prs19}, we were able to capture the dominant torus corrections to the infinite plane results. In this paper, we complete this work by considering the torus corrections of three- and four-point connectivities, which are expected to test much more non-trivial structure constants. Indeed these corrections contain structure constants which do not satisfy any differential equations, contrary to the two-point case.

\noindent In Section \ref{sec:gen} we give the framework within which our problem is stated. In Section \ref{CFTapproach} we outline the CFT approach to study universal finite-size effects on the torus. In Section \ref{sec:stateofart} we define the $Q-$ Potts model and the lattice observables we will consider and we review the relevant results about the CFT describing the critical $Q$-Potts model. In Section \ref{sec:newresults} we give the new theoretical predictions about the three- and four-point connectivities. These predictions are compared to Monte Carlo simulations. In Section \ref{sec:conclusions} we summarise and discuss the results.

\section{Scaling limit of multi-point observables on double-periodic lattices}
\label{sec:gen}
\begin{figure}
	\begin{tikzpicture}[scale=0.4]
		\begin{scope}[on background layer]
			\node (A) at (0.4, -0.4) {};
			\node (B) at (0.4,15.45) {};
			\node (Ap) at(-0.4,0.4) {};
			\node (C) at (10.4,0.4) {};
			 \draw[thick,<->](-0.8,0)--(-0.8,15);	
            \draw[thick,<->](0,-0.2)--(10,-0.2);		
			\draw (4.5,-1) node{$N'$};
			\draw (-1.4,7) node{$M'$};
			\draw(5,15) node[above]{Periodic in both directions};		
			\foreach \x in {0,0.8,...,10} {
				\draw [gray](0.2+\x,0) -- (0.2+\x,15);
				}
			\foreach \y in {0,0.8,...,15} {
				\draw [gray](0,\y+0.2) -- (10,\y+0.2);
				}
			
		\end{scope}
		\filldraw[red] (3.4,3.4) circle (3pt);
			\node (P1) at (2.4, 3.4) {$w_1$};
			\filldraw[red] (7.4,3.4) circle (3pt);
			\node (P2) at (8.4, 3.4) {$w_2$};
			\filldraw[red] (3.4,13) circle (3pt);
			\node (P3) at (2.4, 13) {$w_3$};
			\filldraw[red] (7.4,13) circle (3pt);
			\node (P4) at (8.4, 13) {$w_4$};
			\draw[thick,red](3.4,3.4)--(3.4,13)--(7.4,13)--(7.4,3.4)--(3.4,3.4);
			\draw[thick,red](3.4,13)--(7.4,3.4);
		\draw (15,7.5) node[above]{$M\gg M', N\gg N'$};
		\draw(15,6)node[above]{$\frac{M'}{N'}=\frac{M}{N}$};
		\draw (15,5) node[below]{$|w_i-w_j| \gg 1$};
		\draw[thick,->](12,10.25)--(18,10.25); 
		\begin{scope}[xshift=20cm]
		\filldraw[red] (3.4,3.4) circle (3pt);
			\node (P1) at (2.4, 3.4) {$w_1$};
			\filldraw[red] (7.4,3.4) circle (3pt);
			\node (P2) at (8.4, 3.4) {$w_2$};
			\filldraw[red] (3.4,13) circle (3pt);
			\node (P3) at (2.4, 13) {$w_3$};
			\filldraw[red] (7.4,13) circle (3pt);
			\node (P4) at (8.4, 13) {$w_4$};
			\draw[thick,red](3.4,3.4)--(3.4,13)--(7.4,13)--(7.4,3.4)--(3.4,3.4);
			\draw[thick,red](3.4,13)--(7.4,3.4);
		\end{scope}
		\begin{scope}[xshift=20cm,on background layer]
			\node (A) at (0.4, -0.4) {};
			\node (B) at (0.4,15.45) {};
			\node (Ap) at(-0.4,0.4) {};
			\node (C) at (10.4,0.4) {};
			\draw[thick,<->](10.2,0)--(10.2,15);	
            \draw[thick,<->](0,-0.2)--(10,-0.2);	
            \draw (11,7) node{$M$};
			\draw(5,15) node[above]{Periodic in both directions};	
			\foreach \x in {0,0.2,...,10} {
				\draw[gray] (0.2+\x,0) -- (0.2+\x,15);
				}
			\foreach \y in {0,0.2,...,15} {
				\draw[gray] (0,\y+0.2) -- (10,\y+0.2);
				}
		\end{scope}
	\end{tikzpicture}
	\caption{Scaling limit of multi-point observables on double-periodic lattices.}
	\label{Fig1}
	\end{figure}
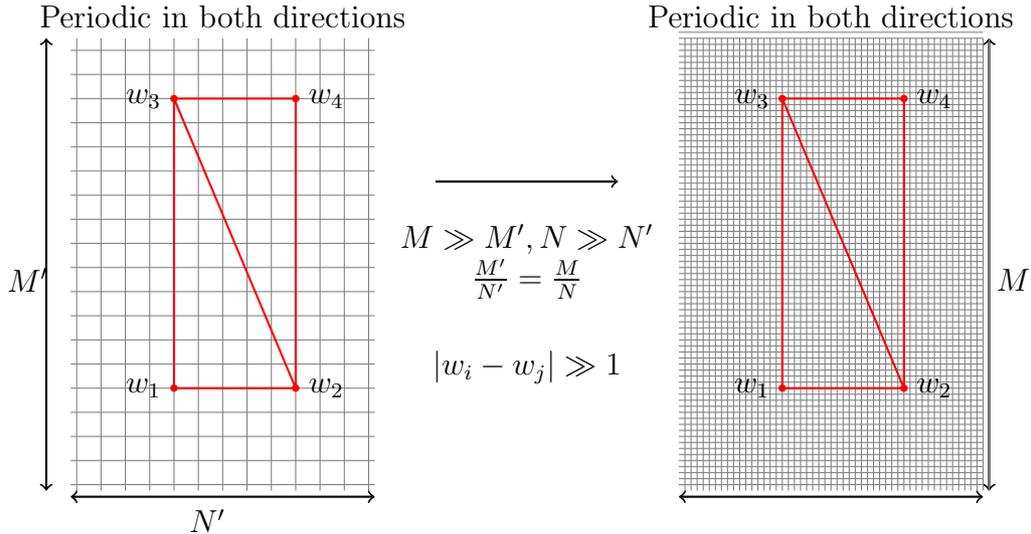

Consider a lattice statistical model that undergoes a second-order phase transition, and define it on  a  $M\times N$ square lattice of mesh $a_0=1$ with double-periodic boundary conditions. The lattice has then the topology of a torus with nome $q$: 
\begin{equation}
\label{toruspar}
q= e^{2 \pi i \tau}, \quad \tau=i \frac{M}{N}.
\end{equation}  
To characterise the universality class, one defines a lattice observable $\mathcal{O}(w_1,\cdots,w_n)$, with $w_1,\cdots,w_n$ indicating points on the torus, and studies its scaling limit $P(w_1,\cdots,w_n)$ at the critical point,  see Fig. \ref{Fig1}.  Supposing that $\mathcal{O}$ is multiplicatively renormalisable (see Chapter 2 in \cite{cardy2008conformal}) one has:
\begin{equation}
P(w_1,w_2,\cdots,w_n)=\lim_{\substack{N\to \infty \\ \frac{M}{N}=O(1)}}N^{2 n\Delta}\;\mathcal{O}(w_1,\cdots,w_n)
\end{equation}
where $2\Delta$ is the scaling dimension associated to the lattice observable. One may think for instance of the $n-$ point Ising spin correlation function at the ferromagnetic-paramagnetic transition. In this case the scaling dimension is $2\Delta=\frac18$, as rigorously proven in \cite{Chelkak12}.     
\noindent \noindent The basic assumptions we will work with are:
\begin{itemize}
\item The system is conformal invariant.
\item When $w_i-w_j\gg1$, $P(w_1,\cdots,w_n)$ is given by the torus $n$-point correlation function of spinless primary fields  with scaling dimension $2\Delta$. 
\item The corresponding CFT has a discrete spectrum.
\end{itemize} 
\noindent The limit \begin{equation}
\frac{|w_i-w_j|}{N}\to 0
\end{equation}
corresponds to the infinite plane limit. In the $n=2,3$ case, the conformal invariance fixes the spatial dependence:
 \begin{equation}
P(w_1,w_2)\xrightarrow{\frac{w_{12}}{N}\to 0} \frac{c_0^{(2)}}{|w_{12}|^{4\Delta}},\quad P(w_1,w_2,w_3)\xrightarrow{\frac{w_{ij}}{N}\to 0} c^{(3)}_{0} \frac{D}{|w_{12}w_{13}w_{23}|^{2\Delta}},
\end{equation}
where $D$ is an universal constant. For $n=4$:  
\begin{equation}
\label{eq:crossratio}
P(w_1,w_2,w_3,w_4)\xrightarrow{\frac{w_{ij}}{N}\to 0}  \frac{c^{(4)}_{0}}{|w_{12}w_{34}|^{4\Delta}}\;P\left(z\right), \quad z= \frac{w_{12}w_{34}}{w_{13}w_{24}},
\end{equation} which means that the problem has been reduced to the computation of a function $P(z)$, with $z$ the cross-ratio.  The $c_{0}^{(n)}$ in the above expressions are non-universal constants. 
\noindent In this paper we will study the behaviour  
of $P(w_1,\cdots,w_n)$ when the distances between points
\begin{equation}
0<\frac{|w_i-w_j|}{N}\ll 1,
\end{equation}
are small but different from zero. In this case we expect corrections to the infinite plane limit coming from the torus topology:    
\begin{align} 
&P(w_1,w_2)=\frac{c^{(2)}_0}{|w_{12}|^{4\Delta}}\left[1+ f^{(2)}_{\tau}\left(\frac{w_{12}}{N}\right)\right]\label{def:f2}\\
&P(w_1,w_2,w_3)=\frac{c_0^{(3)}}{|w_{12}|^{4\Delta} |w_{23}|^{2\Delta}}\left[D\;\left|\frac{w_{12}}{w_{23}}\right|^{2\Delta}\left|\left(1+\frac{w_{12}}{w_{23}}\right)\right|^{-2\Delta}+f^{(3)}_{\tau}\left(\frac{w_{12}}{w_{23}},\frac{w_{23}}{N}\right)\right]\label{def:f3}\\
&P(w_1,w_2,w_3,w_4)=\frac{c_0^{(4)}}{|w_{12}|^{4\Delta}|w_{34}|^{4\Delta}}\left[P\left(\frac{w_{12}w_{34}}{w_{13}w_{24}}\right)+ f^{(4)}_{\tau}\left(\frac{w_{12}w_{34}}{w_{13}w_{24}},\frac{w_{24}}{N}\right)\right].\label{def:f4}
\end{align}
 The functions $f_{\tau}$, symmetric under the replacements $N\leftrightarrow M, \tau \leftrightarrow -\tau^{-1}$, encode the corrections to the infinite plane limit.
\noindent   The assumption that $P(w_1,\cdots,w_n)$ is given by a correlator of local fields in some CFT can be considered quite optimistic if it is applied to non-local observables, such as the geometric properties of critical fractals. Actually we will study these types of observables, namely the $n-$ point connectivities of critical Potts clusters.  However, we will show that the CFT approach not only well describes the plane limit of the cluster connectivities  \cite{prs16,prs19} but also captures the very non-trivial effects of the lattice topology.  
\noindent The functions $f_{\tau}$ are known only in a few cases, namely when the CFT is the compactified free boson  \cite{fsz87Torus,SaleurAttorus}, as in the case of the Ising energy and spin correlation functions, or when a Coulomb gas description is available \cite{Bagger89,narain90}. However, as we will show below, these functions can always be expressed as multiple series expansions. This approach is useful for lattice sizes $N\gg 1$ and location of points $\{w_{i}\}$ for which the series converge quickly. 
\section{Conformal Field Theory approach}
\label{CFTapproach}
\noindent  We outline here how to compute the large $N$ expansion of the functions $f_{\tau}$ in (\ref{def:f2})-(\ref{def:f4}). 
 Let us consider a CFT with: 
 \begin{equation}
 \text{central charge:}\quad  c,\quad \text{and spectrum:}\quad \mathcal{S}
 \end{equation}
defined on the torus (\ref{toruspar}). The central charge $c$ is the parameter that defines the algebra of the conformal generators, the Virasoro algebra (\ref{Vir}). The set of the Virasoro representations entering a  CFT forms its spectrum $\mathcal{S}$.  We refer the reader to \cite{rib14} for an introduction to CFT. Henceforth we indicate $V_{(\Delta)}$ a primary field with $(\Delta)=\Delta,\bar{\Delta}$ its left and right dimensions. The notation $(\Delta)$ will refer either to the highest weight state associated to $V_{(\Delta)}$  or to the entire representation formed by the set of descendants states. The symbol $(\Delta,Y)=\Delta,Y, \bar{\Delta},\bar{Y}$ denotes one of the descendant states with dimensions $\Delta+|Y|$ and $\bar{\Delta}+ |\bar{Y}|$, where $|Y|$, $|\bar{Y}|\in \mathbb{N}$ are the levels of the descendant: as reviewed in  Appendix \ref{sec:defs}, this notation comes from the fact that the descendant states forming a basis of an irreducible representation are labelled by the Young tableaux $Y,\bar{Y}$ with number of boxes $|Y|$ and $|\bar{Y}|$. 

\noindent A CFT is solved when, in addition to the central charge and the spectrum, the structure constants $D_{(\Delta_1),(\Delta_2)}^{(\Delta_3)}$, defined in (\ref{OPE}) and in (\ref{def:D}), are known.

\noindent Suppose we consider a case where the CFT is solved. In particular we are interested in the functions:
 \begin{equation}
\langle V_{(\Delta_1)}(w_1)\cdots V_{(\Delta_n)}(w_n)\rangle,\quad n=2,3,4
 \end{equation}
where $\left<\cdots \right>$ denotes the CFT correlation on the torus (\ref{toruspar}). Notice that $V_{(\Delta)}$ may not be uniquely defined by its scaling dimensions. This is the case, for instance, when the CFT has an  additional symmetry with multiplicities in the representations. 
 
 \subsection{CFT partition function}
The CFT partition function $Z$ takes the form:
 \begin{equation}
Z= \sum_{(\Delta)\in \mathcal{S}} q^{\Delta-c/24}\bar{q}^{\bar{\Delta}-c/24} \left[1+O(q,\bar{q})\right],\label{partition}
\end{equation}
where $q$ is the nome given in (\ref{toruspar}) and the terms in the square brackets correspond to all the contributions of order $O(q^{|Y|}\bar{q}^{|\bar{Y}|})$, coming from the descendants states $(\Delta,Y)$.

 \subsection{One-point function}
\noindent  The one-point function $\langle V_{(\Delta,Y)}\rangle$ on a torus of size $N$ has the expression \cite{fl09}:
\begin{align}
\label{eq:1-pt}
\left< V_{(\Delta,Y)}\right> &=\frac{1}{Z}\frac{(2\pi)^{\Delta+|Y|+\bar{\Delta}+|\bar{Y}|}}{N^{\Delta+\bar{\Delta}+|Y|+|\bar{Y}|}}\sum_{(\Delta_{\text{top}})\in \mathcal{S}}D_{(\Delta),(\Delta_{\text{top}})}^{(\Delta_{\text{top}})}q^{\Delta_{\text{top}}-c/24}\bar{q}^{\bar{\Delta}_{\text{top}}-c/24}\left[1+O(q,\bar{q})\right]\nonumber \\
&=\frac{1}{N^{\Delta+\bar{\Delta}+|Y|+|\bar{Y}|}}\left< V_{(\Delta,Y)}\right>_{(N=1)},
\end{align}
where in order to make the dependence on $N$ more explicit, we introduced, as in \cite{jps19two}, the notation $ \left< \cdots\right>_{(N=1)}$ to indicate a CFT correlation computed on the torus (\ref{toruspar}) with $N=1$. 

\noindent  The representations $(\Delta_{\text{top}})$ contributing to the one-point $V_{(\Delta,Y)}$ torus function are the ones for which the structure constant $D_{(\Delta),(\Delta_{\text{top}})}^{(\Delta_{\text{top}})}$ does not vanish, and which satisfy the fusion rule  $(\Delta_{\text{top}})\times (\Delta_{\text{top}})\to (\Delta)$. Each term appearing in the sum (\ref{eq:1-pt}) is given in (\ref{app1pt}) and can be represented by the diagram:
\begin{figure}[H]
 \centering
 \begin{tikzpicture}[baseline=(current  bounding  box.center), very thick, scale = .6]
 \begin{scope}[xshift = -10cm]
\draw (0,0) -- (0,3);
\draw (0,4.8) circle(1.8cm);
\draw (0,6.5) node [above] {$V_{(\Delta_{\text{top}},Y_{\text{top}})}$};
\draw (0,1.5) node [right] {$V_{(\Delta,Y)}$};\end{scope}
\draw(-3,1.5) node[left]{$=$};
\draw(-3,1.5) node[right]{$D_{(\Delta_{\text{top}},Y_{\text{top}}),(\Delta_{\text{top}},Y_{\text{top}})}^{(\Delta,Y)}q^{\Delta_{\text{top}}+|Y_{\text{top}}|-c/24}\bar{q}^{\bar{\Delta}_{\text{top}}+|\bar{Y}_{\text{top}}|-c/24}$};
\end{tikzpicture}
\caption{Diagrammatic representation of the torus one-point function.}
\label{fig:1pt}
\end{figure}
\noindent where we denote $D_{(\Delta_1,Y_1),(\Delta_2,Y_2)}^{(\Delta_3,Y_3)}$ the constant associated to the three-point function of descendant fields, which is directly proportional to $D_{(\Delta_1),(\Delta_2)}^{(\Delta_3)}$, see (\ref{Dfact}).

\subsection{Two-point function}
\noindent The two-point function $\langle V_{(\Delta_1)}(w_1)V_{(\Delta_2)}(w_2)\rangle$ can be represented in the form:
\begin{align}\label{2pt}
\langle V_{(\Delta_1)}(w_1)V_{(\Delta_2)}(w_2)\rangle = &\frac{1}{\left|w_{12}\right|^{2\Delta_1+2\Delta_2}}\sum_{(\Delta_{\text{top}})\in \mathcal{S}} D_{(\Delta_1),(\Delta_2)}^{(\Delta_{\text{top}})}\left(\frac{w_{12}}{N}\right)^{\Delta_{\text{top}}}\left(\frac{\bar{w}_{12}}{N}\right)^{\bar{\Delta}_{\text{top}}}\left[\left< V_{(\Delta_{\text{top}})}\right>_{(N=1)} +\right.\nonumber \\
&\left.+O\left(\frac{w_{12}}{N},\frac{\bar{w}_{12}}{N}\right)\right].
\end{align}
\noindent 
The contributions in the square bracket come from the descendants $V_{(\Delta_{\text{top}},Y_{\text{top}})}$. The $1/N$ scaling of the topological corrections is then determined by the fields $V_{(\Delta_{\text{top}},Y_{\text{top}})}$. Each of these terms is given in Appendix \ref{subsec:12pt} and is associated to the diagram:
\begin{figure}[H]
 \centering
 \begin{tikzpicture}[baseline=(current  bounding  box.center), very thick, scale = .6]
\draw (-5,0) node [left] {$V_{(\Delta_1)}(w_1)$}--(-2,0)--(1,0);
\draw (-2,0) -- (-2,3);
\draw (-2,4.8) circle(1.8cm);
\draw (-2,1.8) node [right] {$V_{(\Delta_{\text{top}},Y_{\text{top}})}$};
\draw (1,0) node [right] {$V_{(\Delta_2)}(w_2)$};
\draw(4,1.8) node[left]{$=$};
\draw(4,1.8) node[right]{$D_{(\Delta_1),(\Delta_2)}^{(\Delta_{\text{top}},Y_{\text{top}})}\left(\frac{w_{12}}{N}\right)^{\Delta_{\text{top}}+|Y_{\text{top}}|}\left(\frac{\bar{w}_{12}}{N}\right)^{\bar{\Delta}_{\text{top}}+|\bar{Y}_{\text{top}}|}\left<V_{(\Delta_{\text{top}},Y_{\text{top}})}\right>_{(N=1)}.$};
\end{tikzpicture}
\caption{Diagrammatic representation of the torus two-point function.}
\label{fig:2pt}
\end{figure} 
\noindent When the field $V_{(\Delta_{\text{top}},Y_{\text{top}})}=\text{Id}$ is the identity field, i.e. $\Delta_{\text{top}}=\bar{\Delta}_{\text{top}}=0, |Y_{\text{top}}|=|\bar{Y}_{\text{top}}|=0$ and with $\left<\text{Id}\right>=1$, one recovers the plane limit (the primary fields are normalised such that $D_{(\Delta_1),(\Delta_2)}^{\text{Id}}=1$). Setting $(\Delta_1)=(\Delta_2)=(\Delta)$, we find the expansion of $f^{(2)}_{\tau}(\frac{w_{12}}{N})$ in (\ref{def:f2}):
\begin{equation}
\label{eq:f2exp}
f^{(2)}_{\tau}\left(\frac{w_{12}}{N}\right)=D_{(\Delta),(\Delta)}^{(\Delta)_{\text{min}}}\left(\frac{w_{12}}{N}\right)^{\Delta_{\text{min}}}\left(\frac{\bar{w}_{12}}{N}\right)^{\bar{\Delta}_{\text{min}}}\left<V_{(\Delta)_{\text{min}}}\right>_{(N=1)}+o\left(\frac{1}{N^{\Delta_{\text{min}}+\bar{\Delta}_{\text{min}}}}\right),
\end{equation}
where $V_{(\Delta)_{\text{min}}}$ is the state among the $V_{(\Delta_{\text{top}})}$ appearing in the $(\Delta)\otimes(\Delta)$ fusion with lowest dimensions $\Delta_{\text{min}},\bar{\Delta}_{\text{min}}$.  Note that the assumption made in Section \ref{sec:gen} of the discreteness of the spectrum $\mathcal{S}$ implies that the dimensions of the fields are discretely spaced. A more detailed treatment of the two-point function can be found in \cite{jps19two}.
\subsection{Three-point function}
\noindent 
In the channel expansion where $w_1\to w_2$, $\langle V_{(\Delta_1)}(w_1)V_{(\Delta_2)}(w_2)V_{(\Delta_3)}(w_3)\rangle$ can be expressed as:
 \begin{align}
\left< V_{(\Delta_1)}(w_1)V_{(\Delta_2)}(w_2)V_{(\Delta_3)}(w_3)\right>=&\left|w_{12}\right|^{-2\Delta_1-2\Delta_2}\sum_{(\Delta_L)\in \mathcal{S}}D_{(\Delta_1),(\Delta_2)}^{(\Delta_L)}w_{12}^{\Delta_L}\bar{w}_{12}^{\bar{\Delta}_L}\Big[\left< V_{(\Delta_L)}(w_2)V_{(\Delta_3)}(w_3)\right>\nonumber \\
&+O\left(w_{12},\bar{w}_{12}\right)\Big].
\end{align}
The contributions in the square brackets come from the descendants of $V_{(\Delta)_L}$ and are given in Appendix \ref{sec:3ptder}.  Expanding the two-point function, similarly to what has been done above, one finds that each of these corrections can be associated to the diagram: 
\begin{figure}[H]
 \centering
 \begin{tikzpicture}[baseline=(current  bounding  box.center), very thick, scale = .6]
\draw (-3,0) node [left] {$V_{(\Delta_1)}$}--(0,0)--(0,3) node[above]{$V_{(\Delta_2)}$};
\draw (0,0)  -- (8,0);
\draw (2.,0) node[below]{$V_{(\Delta_L,Y_L)}$};
\draw (4,0) -- (4,2.5);
\draw (4,4.3) circle(1.8cm);
\draw (4,1.5) node [right] {$V_{(\Delta_{\text{top}},Y_{\text{top}})}$};
\draw (8,0) node [right] {$V_{(\Delta_3)}$};
\draw (-5,-2.5) node [left] {$=$};
\draw (-5,-2.5) node [right] {$D_{(\Delta_1),(\Delta_2)}^{(\Delta_L,Y_L)}D_{(\Delta_L,Y_L),(\Delta_3)}^{(\Delta_{\text{top}})}\; \left(\frac{w_{12}}{w_{23}}\right)^{\Delta_L+|Y_L|}\left(\frac{\bar{w}_{12}}{\bar{w}_{23}}\right)^{\bar{\Delta}_L+|\bar{Y}_L|}\left(\frac{w_{23}}{N}\right)^{\Delta_{\text{top}}+|Y_{\text{top}}|}\left(\frac{\bar{w}_{23}}{N}\right)^{\bar{\Delta}_{\text{top}}+|\bar{Y}_{\text{top}}|} \left< V_{(\Delta_{\text{top}},Y_{\text{top}})}\right>$};
\end{tikzpicture}
\caption{Diagrammatic representation of the torus three-point function.}
\label{fig:3pt}
\end{figure}
\noindent The terms with $\Delta_{\text{top}}=\bar{\Delta}_{\text{top}}=0, |Y_{\text{top}}|=|\bar{Y}_{\text{top}}|=0$ add up to give the plane limit. 
We can specify now to the case $(\Delta_1)=(\Delta_2)=(\Delta_3)=(\Delta)$ and give the form of the double expansion of $f_{\tau}^{(3)}$ in (\ref{def:f3}):
\begin{align}
\label{eq:f3fin}
&f_{\tau}^{(3)}\left(\frac{w_{12}}{w_{23}},\frac{w_{23}}{N}\right)=c^{(3)}_{\text{min}}\left(\frac{w_{12}}{w_{23}},\tau\right) \left(\frac{w_{23}}{N}\right)^{\Delta_{\text{min}}}\left(\frac{\bar{w}_{23}}{N}\right)^{\bar{\Delta}_{\text{min}}} +o\left(\frac{1}{N^{\Delta_{\text{min}}+\bar{\Delta}_{\text{min}}}}\right).
\end{align}
The coefficient $c^{(3)}_{\text{min}}$ is given by: 
\begin{align}\label{defc3}
&c^{(3)}_{\text{min}}\left(\frac{w_{12}}{w_{23}},\tau\right)= \left< V_{(\Delta_{\text{min}})}\right> \;\sum_{(\Delta_L,Y_L)\in \mathcal{S}}D_{(\Delta),(\Delta)}^{(\Delta_L,Y_L)}D_{(\Delta_L,Y_L),(\Delta)}^{(\Delta_{\text{min}})}\left(\frac{w_{12}}{w_{23}}\right)^{\Delta_L+|Y_L|}\left(\frac{\bar{w}_{12}}{\bar{w}_{23}}\right)^{\bar{\Delta}_L+|\bar{Y}_L|}.
\end{align}
\noindent The field $V_{(\Delta_{\text{min}})}$ corresponds to the state with lowest dimensions $\Delta_{\text{min}},\bar{\Delta}_{\text{min}}$ appearing in the $(\Delta_L)\otimes(\Delta)$ fusion, and therefore it can be different from the one appearing in (\ref{eq:f2exp}).
 \subsection{Four-point function}
In the $s-$ channel, the four-point function admits the following expansion:
 
\begin{align}
&\left<V_{(\Delta_1)}(w_1)V_{(\Delta_2)}(w_2)V_{(\Delta_3)}(w_3)V_{(\Delta_4)}(w_4)
\right> =\sum_{(\Delta_L),(\Delta_R)\in\mathcal{S}} D_{(\Delta_1),(\Delta_2)}^{(\Delta_L)}D_{(\Delta_3),(\Delta_4)}^{(\Delta_R)}\times \nonumber \\ &\times w_{12}^{\Delta_l-\Delta_1-\Delta_2} \bar{w}_{12}^{\bar{\Delta}_L-\bar{\Delta}_1-\bar{\Delta}_2}w_{34}^{\Delta_R-\Delta_3-\Delta_4} \bar{w}_{34}^{\bar{\Delta}_R-\bar{\Delta}_3-\bar{\Delta}_4}\left[\left<V_{(\Delta_L)}(w_2)V_{(\Delta_R)}(w_4)\right>+\right. \nonumber\\ 
&\left. +O\left(w_{12},\bar{w}_{12}, w_{34},\bar{w}_{34}\right) \right].
\end{align}
As explained in Appendix \ref{sec:4ptder}, each term of the above sum is represented by the diagram :
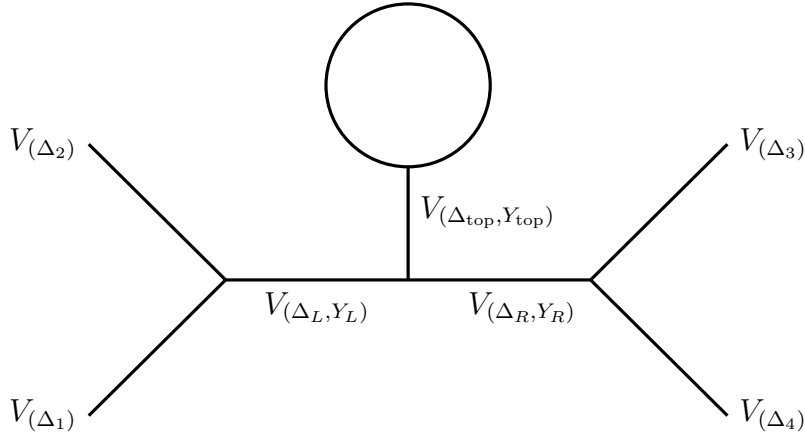
\begin{figure}[H]
 \centering
 \begin{tikzpicture}[baseline=(current  bounding  box.center), very thick, scale = .6]
\draw (-3,-3) node [left] {$V_{(\Delta_1)}$}--(0,0)--(-3,3) node[left]{$V_{(\Delta_2)}$};
\draw (0,0)  -- (8,0);
\draw (2,0) node[below]{$V_{(\Delta_L,Y_L)}$};
\draw (4,0) -- (4,2.5);
\draw (4,4.3) circle(1.8cm);
\draw (4,1.5) node [right]{$V_{(\Delta_{\text{top}},Y_{\text{top}})}$};
\draw (6.5,0) node[below]{$V_{(\Delta_R,Y_R)}$};
\draw (11,3) node [right] {$V_{(\Delta_3)}$}--(8,0)--(11,-3) node[right]{$V_{(\Delta_4)}$};
\draw(-9,-6) node[left]{$=$};
\draw (-9,-6) node[right]{$ D_{(\Delta_1),(\Delta_2)}^{(\Delta_L,Y_L)}D_{(\Delta_3),(\Delta_4)}^{(\Delta_R,Y_R)} D_{(\Delta_L,Y_L),(\Delta_R,Y_R)}^{(\Delta_{\text{top}})}\left(\frac{w_{12}}{w_{24}}\right)^{\Delta_L+|Y_L|} \left(\frac{\bar{w}_{12}}{\bar{w_{24}}}\right)^{\bar{\Delta}_l+|\bar{Y}_L|}\left(\frac{w_{34}}{w_{24}}\right)^{\Delta_L+|Y_L|} \left(\frac{\bar{w}_{34}}{\bar{w}_{24}}\right)^{\bar{\Delta}_R+|\bar{Y}_R|}$};
\draw(-7,-8) node[right]{$\times \left(\frac{w_{24}}{N}\right)^{\Delta_{\text{top}}+|Y_{\text{top}}|}\left(\frac{\bar{w}_{24}}{N}\right)^{\bar{\Delta}_{\text{top}}+|\bar{Y}_{\text{top}}|}\left<V_{(\Delta_{\text{top}},Y_{\text{top}})}\right>$};
\end{tikzpicture}\caption{Diagrammatic representation of the torus four-point function.}
\label{fig:4pt}
\end{figure}
\noindent  The sum over the diagrams with $\Delta_{\text{top}}=\bar{\Delta}_{\text{top}}=0, |Y_{\text{top}}|=|\bar{Y}_{\text{top}}|=0$ coincides with the $s-$ channel expansion of the plane four-point correlation function \cite{rib14}. Setting $(\Delta_i)=(\Delta)$, $i=1,\cdots 4$, the multi-series expansion of $f^{(4)}_{\tau}$ in (\ref{def:f4}) takes the form:
\begin{align}
\label{eq:f4fin}
&f^{(4)}_{\tau}\left(\frac{w_{12}}{w_{24}},\frac{w_{34}}{w_{24}},\frac{w_{24}}{N}\right)
=c^{(4)}_{\text{min}}\left(\frac{w_{12}}{w_{24}},\frac{w_{34}}{w_{24}},\tau\right) \left(\frac{w_{24}}{N}\right)^{\Delta_{\text{min}}}\left(\frac{\bar{w}_{24}}{N}\right)^{\bar{\Delta}_{\text{min}}} +o\left(\frac{1}{N^{\Delta_{\text{min}}+\bar{\Delta}_{\text{min}}}}\right),
\end{align}
where:
\begin{align}\label{defc4}
c^{(4)}_{\text{min}}\left(\frac{w_{12}}{w_{24}},\frac{w_{34}}{w_{24}},\tau\right)=& \left< V_{(\Delta_{\text{min}})}\right> \;\sum_{\substack{(\Delta_L,Y_L),\\(\Delta_R,Y_R)\in \mathcal{S}}}D_{(\Delta),(\Delta)}^{(\Delta_L,Y_L)}D_{(\Delta),(\Delta)}^{(\Delta_R,Y_R)}D_{(\Delta_L,Y_L),(\Delta_R,Y_R)}^{(\Delta_{\text{min}})}\nonumber \\
&\times \left(\frac{w_{12}}{w_{24}}\right)^{\Delta_L+|Y_L|} \left(\frac{\bar{w}_{12}}{\bar{w_{24}}}\right)^{\bar{\Delta}_L+|\bar{Y}_L|}\left(\frac{w_{34}}{w_{24}}\right)^{\Delta_R+|Y_R|} \left(\frac{\bar{w}_{34}}{\bar{w}_{24}}\right)^{\bar{\Delta}_R+|\bar{Y}_R|}.
\end{align}
\section{$Q-$ Potts random cluster  model: CFT versus Monte Carlo results}
\label{sec:stateofart}
We want to apply the above formulas to the study of the connectivities of the  $Q-$ Potts clusters. 
\subsection{Lattice model and multi-point connectivities} 
Let us consider a double-periodic square lattice with parameters (\ref{toruspar}) whose edges can carry a bond or not. The random cluster $Q$-state Potts model \cite{FK72} on such a lattice is defined by the partition function 
\begin{align}
 \mathcal{Z}_Q=\sum_{\mathcal{G}} Q^{\#\,\mathrm{clusters}} p^{\#\, \mathrm{bonds}} (1-p)^{\#\, \text{edges without bond}},
 \label{eq:z}
\end{align}
where $\mathcal{G}$ denotes one of the possible bond configurations and $p\in [0,1]$.
The clusters percolate at the critical value
\begin{align}
\label{criticalp}
p=p_c=\frac{\sqrt{Q}}{\sqrt{Q}+1},
\end{align}
 The lattice multi-point observables $\mathcal{O}$ at $p=p_c$ we consider is: 
\begin{align}
&\mathcal{O}(w_1,\cdots,w_n)=\text{Probability}(w_1,\, w_2,\,\cdots, w_n \text{ are in the same cluster}). 
\end{align}
For $n=2,3$ the above probabilities scan the space of all possible connectivities, while, for $n=4$, the space of connectivities is four dimensional \cite{DeViconn}, see also \cite{js18}-\cite{prs19}. Here we will focus only on the above type of connectivity.

\subsection{The CFT describing the critical Potts cluster model: state of the art.}
\noindent We parametrise the central charge $c$ and the conformal dimension $\Delta$ as follows: 
\begin{equation}
\label{paradelta}
 c = 1-6\left(\beta -\beta^{-1}\right)^2, \quad \Delta= \Delta_{(r,s)}  = \frac{c-1}{24} + \frac14 \left(r\beta -\frac{s}{\beta}\right)^2.
\end{equation}
A representation is degenerate  if  $r,s \in \mathbb{N^*}$, and has a null state at level $r s$. 
The symbols 
\begin{align}
V_{\Delta_{(r,s)},\Delta_{(r,s)}}=V_{(r,s)^D},\quad V_{\Delta_{(r,s)},\Delta_{(r,-s)}}=V_{(r,s)}
\end{align}
indicate the diagonal and non-diagonal primary fields.  
The notations   
\begin{align}
(r,s)^D, \quad (r,s)
\end{align}
 denote the representations associated to $V_{(r,s)^D}$ and $V_{(r,s)}$ respectively. This allows us to use a lighter notation for the structure constants, for instance:
 \begin{align}
  D_{(r_1,s_1),(r_2,s_2)}^{(r,s)^D}=D^{(\Delta_{r_3,s_3},\Delta_{r_3,s_3})}_{\left( \Delta_{r_1,s_1},\Delta_{r_1,-s_1}\right),\left( \Delta_{r_2,s_2},\Delta_{r_2,-s_2}\right)}.
  \end{align}
\noindent A set of these representations is denoted as 
\begin{align}
\mathcal{S}^{D}_{X}=\{(r,s)^D\}_{(r,s)\in X}, \quad  \mathcal{S}_{X}=\{(r,s)\}_{(r,s)\in X},
\end{align}
where $X$ is a given set of pairs $(r,s)$. A third set type is $\mathcal{S}^{\mathrm{quot}}_X$ that contains the degenerate representations with vanishing null state.

\noindent What do we know about the CFT describing the critical Potts clusters ?  We know 
\begin{itemize}
\item the central charge $c(\beta)$. In the $\beta$ parametrisation (\ref{paradelta}), the critical $Q-$ Potts model is related to a CFT with:
\begin{align}
 Q = 4\cos^2\pi \beta^2 \quad \text{with} \quad \frac12 \leq  \beta^2 \leq 1. 
\label{cq}
\end{align} 
\item the spectrum $\mathcal{S}=\mathcal{S}^{\mathrm{Potts}}$ \cite{fsz87}
 \begin{equation}
\label{sppotts}
\mathcal{S}^{\mathrm{Potts}}=\mathcal{S}^{D,\mathrm{quot}}_{(1,\mathbb{N^*})}\bigcup_{\substack{j\geq 2 \\ M | j, p\wedge M=1}} \mathcal{S}_{(j,\mathbb{Z}+\frac{p}{M})} \bigcup \;\mathcal{S}_{(0,\mathbb{Z}+\frac12)}.
\end{equation}
The multiplicities associated to the above sectors have also been computed \cite{fsz87} and, for general $Q$, assume general real values.
We refer the reader to \cite{js18} and references therein for a detailed  discussion of (\ref{sppotts}). 
\end{itemize}
\noindent  We do not know:
\begin{itemize}
\item the CFT Potts model structure constants.
\noindent In other words, for general $Q$, a complete bootstrap solution of the Potts CFT has not been found yet.
\end{itemize}
\noindent The informations on the central charge and on the spectrum allow the computation of certain probabilities on the torus, such as the cluster wrapping probability \cite{langlands92,pinson94,blanchard2014}, as well as the determination 
of different critical exponents or equivalently of the plane two-point functions. Using a Coulomb gas technique \cite{henkel2012conformal2}, the scaling limit $p_{12}(w_1,w_2)$ of $\mathcal{O}(w_1,w_2)$ is obtained as
\begin{equation}
\label{eq:n-ptscal}
p_{12}(w_1,w_2)=\lim_{\substack{N\to\infty\\N/M=O(1)}}\;N^{4\Delta_{(0,\frac12)}}\;\mathcal{O}(w_1,w_2),
\end{equation}
and the plane limit
\begin{equation}
\label{eq:2-ptplane}
p_{12}(w_1,w_2)\xrightarrow{\frac{w_{12}}{N}\to 0}= \frac{c_0}{|w_1-w_2|^{4\Delta_{(0,\frac12)}}},
\end{equation}
is given by the two-point function of fields $V_{(0,\frac12)}$, for this reason called the connectivity fields. This result implies also that the cluster fractal dimension is $2-2\Delta_{(0,\frac12)}$. In (\ref{eq:2-ptplane}), $c_0$ corresponds to the non-universal constant $c_{0}^{(2)}$ appearing in (\ref{def:f2}) and has been computed numerically in \cite{prs16,prs19,jps19two}. 
In \cite{jps19two}, we set:
\begin{equation}
\label{asstwo} 
p_{12}(w_1,w_2)=c_0 \left<V_{(0,\frac12)}(w_1)V_{(0,\frac12)}(w_2)\right>,
\end{equation}
and we made the assumption that the connectivity fields entering the two-point function admit the following fusion
\begin{equation}
\label{eq:OPE2pt}
V_{(0,\frac12)}\otimes V_{(0,\frac12)}=V_{(1,1)^D}\oplus V_{(1,2)^D}\oplus V_{(1,3)^D}+\cdots
\end{equation}
The representations in $\mathcal{S}^{D,\mathrm{quot}}_{(1,\mathbb{N^*})}$ have vanishing null-states, and this fixes their fusion rules \cite{rib14}. For instance, for the field $(1,2)^D$ one has:
\begin{equation}
\label{eq:ope12}
V_{(1,2)^D}\times V_{(r,s)}= V_{(r,s+1)}\oplus V_{(r,s-1)}.
\end{equation} 
Moreover, the structure constants $D^{(1,s)^D}_{(\Delta_1)(\Delta_2)}$ can be exactly computed and expressed in terms of $\Gamma$ functions \cite{rib14}.  In \cite{jps19two} we showed that using (\ref{eq:OPE2pt}), we obtained very good predictions for $p_{12}$.
\noindent Following (\ref{eq:n-ptscal}), we define the scaling limits:
\begin{equation}
\label{def:pn}
p_{12\cdots n}(w_1,w_2,\cdots,w_n)=\lim_{\substack{N\to\infty\\N/M=O(1)}}\;N^{2 n\Delta_{(0,\frac12)}}\;\mathcal{O}(w_1,w_2,\cdots w_n).
\end{equation}
The plane limits of $p_{123}$ and $p_{1234}$ have been at the center of an intense research activity in the last few years as they may directly probe the CFT structure constants. Let us consider the plane limit of $p_{123}$ first. As explained in \cite{ijs15}, if the plane $p_{12}$ can be rewritten in terms of an equivalent local model \cite{henkel2012conformal2}, this is no more true for $p_{123}$ which keeps its non-local nature. Despite this, an important progress was done in \cite{devi11} where the plane limit of $p_{123}$ was conjectured to be given by a CFT three-point  correlator of fields $V_{(0,\frac12)}$:
\begin{align}
\label{eq:3-ptplane}
p_{123}(w_1,w_2,w_3)\xrightarrow{\frac{w_{12}}{N}\to 0} c_{0}^{(3)}\frac{D^{(0,\frac12)}_{(0,\frac12),(0,\frac12)}}{|w_{12}w_{13}w_{24}|^{2\Delta_{(0,\frac12)}}}.
\end{align} 
The conjecture in \cite{devi11} is even stronger, as it proposes a value for the structure constant: 
\begin{equation}
D^{(0,\frac12)}_{(0,\frac12),(0,\frac12)}=\sqrt{2}\;C_{(0,\frac12),(0,\frac12)}^{(0,\frac12)},
\end{equation}
where the $ C_{(\Delta_1),(\Delta_2)}^{(\Delta_3)}$ are the $c\leq 1$ Liouville structure constants \cite{zam05,sch05,kp05a,RiSa14}, defined in Appendix \ref{sec:Liouville}. The factor $\sqrt{2}$ in (\ref{eq:3-ptplane}) originates from a two-fold multiplicity of the theory \cite{devi11,ijs15}. Equation (\ref{eq:3-ptplane}) has been numerically checked in \cite{ziff11,psvd13,ei15,ijs15}. Finally, the non-universal constant $c_{0}^{(3)}$ was verified, for a square lattice, to be strictly related to $c_{0}^{(2)}$ \cite{psvd13}:
 \begin{equation}
c_{0}^{(3)}=(c_{0}^{(2)})^{\frac32}=c_{0}^{\frac32}.
\end{equation}
The above result is consistent with the fact that one can associate a non-universal normalisation $c_{0}^{\frac12}$ to each field $V_{(0,\frac12)}$. 
\noindent In \cite{js18} the full space of $n=4$ connectivities has been considered and the set of representations entering the corresponding  $s-$ channel, i.e. small $z$ see (\ref{eq:crossratio}), determined. For $p_{1234}$ the result is:
\begin{align}
\label{eq:4-ptplane1}
&p_{1234}(w_1,w_2,w_3,w_4)\xrightarrow{\frac{w_{ij}}{N}\to 0} c_0^{2}\;|w_{12}w_{34}|^{-4\Delta_{(0,\frac12)}} \;P_{0}(z)\nonumber \\
&P_{0}(z)=\sum_{\substack{(0,s) \\ s \in \frac{2\mathbb{N}+1}{2}}}\; \left(D^{(0,s)}_{(0\frac12),(0,\frac12)}\right)^2\; |z|^{2\;\Delta_{(0,s)}}\left(1+O(z,\bar{z})\right)\nonumber \\
&+\sum_{\substack{(r,s) \\ r \in 2\mathbb{Z}*, s\in \mathbb{Q}, rs\in 2\mathbb{Z}}}\; \left(D^{(r,s)}_{(0\frac12),(0,\frac12)}\right)^2 \; z^{\;\Delta_{(r,s)}}\bar{z}^{\;\Delta_{(r,-s)}}\left(1+O(z,\bar{z})\right)
\end{align}
In \cite{prs16,prs19} it has been shown that using - whenever they are well defined - the $c\leq1$ Liouville structure constants provides an extremely good approximation to the plane $n=4$ connectivities. The dominant terms for $P_{0}(z)$ are:
\begin{align}
\label{eq:4-ptplane2}
&P_{0}(z)=2\left(C_{(0,\frac12)(0,\frac12)}^{(0,\frac12)}\right)^2|z|^{2\Delta_{(0,\frac12)}}
\left(1+O(z)\right)+\left(D_{(0,\frac12),(0,\frac12)}^{(2,0)}\right)^2 |z|^{2\Delta_{(2,0)}}\left(1+O(z)\right)+\nonumber \\
&+2 \left( C_{(0,\frac{1}{2}),(0,\frac{1}{2})}^{(0,\frac{3}{2})} \right)^2 \left|z\right|^{2\Delta_{(0,\frac{3}{2})}} (1+O(z))+\cdots
\end{align}
The value of $\left(D_{(0,\frac12),(0,\frac12)}^{(2,0)}\right)^2$ has been determined numerically in \cite{prs19} when $Q=1$.
 
\section{The dominant torus corrections to $p_{123}$ and $p_{1234}$}
\label{sec:newresults}
We present here the new results concerning the dominant torus correction of $p_{123}$ and $p_{1234}$, defined in (\ref{def:pn}).
Analogously to what we have done for $p_{12}$ \cite{jps19two}, see (\ref{asstwo}),  we assume that
\begin{align}
&p_{123}=c_0^{\frac32}\;\left<V_{(0,\frac12)}(w_1)V_{(0,\frac12)}(w_2)V_{(0,\frac12)}(w_3)\right>\nonumber \\
&p_{1234}=c_0^{2}\;\left<V_{(0,\frac12)}(w_1)V_{(0,\frac12)}(w_2)V_{(0,\frac12)}(w_3)V_{(0,\frac12)}(w_4)\right>,
\end{align}
and we apply the CFT approach outlined in  Section \ref{CFTapproach} by using the fusions (\ref{eq:OPE2pt}), (\ref{eq:ope12}) and the following one:
\begin{align}
\label{eq:OPE34pt}
&V_{(0,\frac12)}\otimes V_{(0,\frac12)}=\mathop{\oplus}_{s \in \frac{2\mathbb{N}+1}{2}} V_{(0,s)}\mathop{\oplus}_{\substack{(r,s) \\ r \in 2\mathbb{Z}*, s\in \mathbb{Q}, rs\in 2\mathbb{Z}}} V_{(r,s)}
\end{align}
This fusion comes from (\ref{eq:3-ptplane}) and (\ref{eq:4-ptplane1}).
We will use the Liouville constants $C_{(0,\frac12),(0,\frac12)}^{(0,2\mathbb{N}+1)}$ for the connectivity sector. There are no predictions for the other structure constants. Luckily we will in general not need them as they produce sub-dominant diagrams for the values of $Q$, $1\leq Q\leq 4$ considered here. The only exception is the  channel $(2,0)$ which produces a small but visible contribution at $Q=1$. In this case we will use the value for  $D_{(0,\frac12),(0,\frac12)}^{(2,0)}$ found numerically in \cite{prs19}. The fusion (\ref{eq:OPE34pt}) is not in principle contradictory with (\ref{eq:OPE2pt}). Indeed, there may exist different fields  $V_{(0,\frac12)}$ with the same dimension but with different fusion rules. This fact can be well understood for $Q=3$ where the primary fields carry a $Z_3$ charge. In this case there are two fields, $V^{\pm 1}_{(0,\frac12)}$ with $Z_3$ charge $\pm 1$. The fields $V^{+}_{(0,\frac12)}V^{-}_{(0,\frac12)}\to V_{(1,1)} \oplus V_{(1,2)}\cdots$ fuse into the $Z_3$ neutral sector, where one finds the identity $V_{(1,1)}$, while the two fields $V^{+}_{(0,\frac12)}V^{+}_{(0,\frac12)}\to V^{-}_{(0,\frac12)} \oplus V^{-}_{(0,\frac12)}\oplus \cdots $ fuse into the sector of charge $-1$. For general $Q$ however, we do not fully understand how to characterise the primary fields $V_{(0,\frac12)}$ to describe the space of connectivity (see also the discussion in Appendix C.3 of \cite{prs19}). 
\noindent Analogously to the $n=2$ case \cite{jps19two}, we find using (\ref{eq:OPE2pt}), (\ref{eq:ope12}) and (\ref{eq:OPE34pt}), that the dominant topological corrections to $p_{123}$ and $p_{1234}$ are associated to the energy one-point function $ \left<V_{(1,2)^D}\right>$. Comparing with (\ref{eq:f3fin}) and (\ref{eq:f4fin}) we have that $\Delta_{\text{min}}= \Delta_{(1,2)^D}$. We stress the fact that the fusions (\ref{eq:OPE2pt}) and (\ref{eq:OPE34pt}) produce diagrams proportional to $ \left<V_{(0,\frac12)}\right>$ and which would be dominant, as $\Delta_{(0,\frac12)}<\Delta_{(1,2)}\;\forall Q$. However we conjecture that the one-point function of the connectivity field vanishes for any $Q$. If this is easy to show by symmetry argument for $Q=2,3,4$, we could not prove it for general $Q$. The agreement of our results with the Monte Carlo measurements supports this conjecture.
\noindent The computation of the functions $f_{\tau}^{(3)}$ and $f_{\tau}^{(4)}$ in (\ref{def:f3}) and (\ref{def:f4}) is based on two approximations:
\begin{itemize}
\item We compute only the diagrams that contribute to the dominant torus correction, which is sufficient in general for comparison with the numerical data. The only exception are the diagrams proportional to $\left<V_{(1,3)^D}\right>$ which produce a sub-dominant contribution that is visible numerically near $Q=3$. Higher $1/N$ corrections coming from descendant fields could in principle be computed but are expected to give very sub-dominant contributions.
\item For any dominant diagram, we compute the contributions of the descendants at levels one and two. As explained in the next subsection, we expect the contribution of level three to be negligible.
\end{itemize}

\subsection{Three-point connectivity $p_{123}$}\label{sec:3pt}
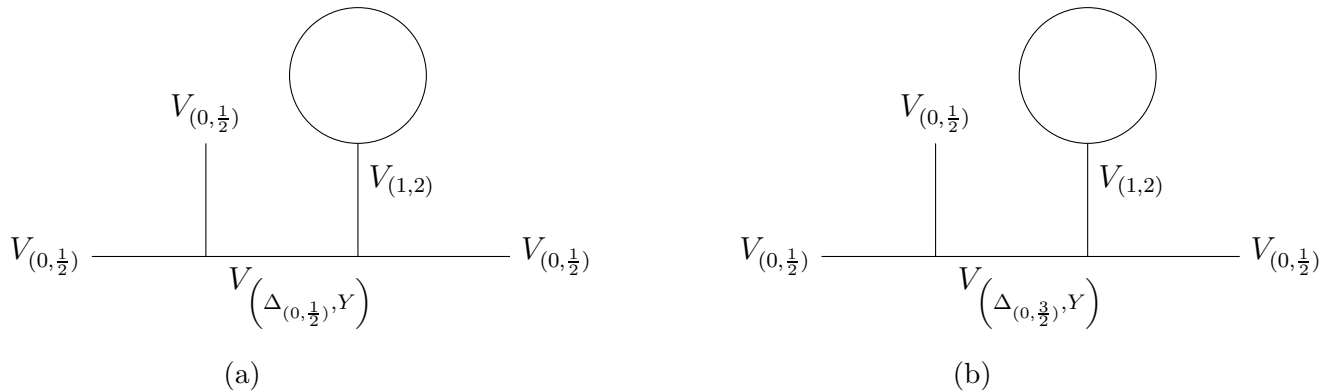
\begin{figure}[H]\centering
\begin{subfigure}{0.4\textwidth}
\begin{tikzpicture}[scale = .5]
\draw (-3,0) node [left] {$V_{(0,\frac12)}$}--(0,0)--(0,3) node[above]{$V_{(0,\frac12)}$};
\draw (0,0)  -- (8,0);
\draw (2.5,0) node[below]{$V_{\left(\Delta_{(0,\frac12)},Y\right)}$};
\draw (4,0) -- (4,3);
\draw (4,4.8) circle(1.8cm);
\draw (4,2) node [right] {$V_{(1,2)}$};
\draw (8,0) node [right] {$V_{(0,\frac12)}$};
\end{tikzpicture}
\caption{ }
\label{fig:channel12}
\end{subfigure}\hfill
\begin{subfigure}{0.4\textwidth}
\begin{tikzpicture}[scale = .5]
\draw (-3,0) node [left] {$V_{(0,\frac12)}$}--(0,0)--(0,3) node[above]{$V_{(0,\frac12)}$};
\draw (0,0)  -- (8,0);
\draw (2.5,0) node[below]{$ V_{\left(\Delta_{(0,\frac32)},Y\right)}$};
\draw (4,0) -- (4,3);
\draw (4,4.8) circle(1.8cm);
\draw (4,2) node [right] {$V_{(1,2)}$};
\draw (8,0) node [right] {$V_{(0,\frac12)}$};
\end{tikzpicture}
\caption{ }
\label{fig:channel32}
\end{subfigure}\caption{Diagrammatic representation of the two channels contributing to the topological corrections of the three-point connectivity.}
\end{figure}

Using the fusion (\ref{eq:ope12}), the only contributions of order $O\left(\left(\frac{w_{23}}{N}\right)^{2\Delta_{(1,2)}}\right)$ come from the fusion channels $V_{(0,\frac12)}\times V_{(0,\frac12)}\to V_{(0,\frac12)}$ and $V_{(0,\frac12)}\times V_{(0,\frac12)}\to V_{(0,\frac32)}$, represented respectively by diagrams \ref{fig:channel12} and \ref{fig:channel32}.

\noindent 
As detailed in Appendix \ref{sec:3ptder} we can compute the coefficient $c^{(3)}_{(1,2)}\left(\frac{w_{12}}{w_{23}},\tau\right)$ defined as:
\begin{equation}\label{p3Potts}
p_{123} = \frac{ c_0^{\frac{3}{2}} }{ \left|w_{12}w_{23}w_{13}\right|^{2\Delta_{(0,\frac{1}{2})}}} \left[ D + \left| \frac{w_{13}}{w_{12}} \right|^{2\Delta_{(0,\frac{1}{2})}} c_{(1,2)}^{(3)}\left(\frac{w_{12}}{w_{23}},\tau\right)  \left|\frac{w_{23}}{N}\right|^{2\Delta_{(1,2)}} + \text{subleading}\right].
 \end{equation}
In particular, to compare with the Monte Carlo numerical data we set $\tau = i$ (i.e. $M=N$),  and we fix the three points $w_1,w_2,w_3$ at the vertices of isosceles triangles. First we consider the configuration $(w_1,w_2,w_3) = (0,r,ir)$, also considered in \cite{psvd13} and compute the coefficient $c^{(3),\text{an}}_{(1,2)}=c^{(3)}_{(1,2)}\left(\frac{1}{\sqrt{2}},\tau=i\right)$ at level $2$, see equation (\ref{c312}) in the Appendix. The comparison with the Monte Carlo results $c^{(3),MC}_{(1,2)}$, for different values of $Q$,  is given in table and figure (\ref{tabfig:iso}). In the figure we show the convergence of our expansion, computed to order $\frac{w_{12}}{w_{23}}$ ie to level one (dashed) and $\left(\frac{w_{12}}{w_{23}}\right)^2$ ie to level two (solid). The contribution of order $\left(\frac{w_{12}}{w_{23}}\right)^3$ is expected to be negligible, below the precision of the numerical measurements.

\begin{figure}[!ht]
    \centering
    \begin{tikzpicture}[baseline=(current  bounding  box.center), very thick, scale = 1.1]
\begin{axis}[
	legend cell align=center,
	xlabel={$Q$},
	ylabel={$c^{(3)}_{\Delta_{(1,2)}}\left(\frac{1}{\sqrt{2}}|\tau=i\right)$},	
	legend pos=south east]
	]
\addlegendentry{Monte Carlo}
\addplot+[green,mark=o,only marks,mark size=.6pt] 
	table [y=MC, x=Q]{tab-levels.dat};
\addlegendentry{level 1}
\addplot[gray,dashed]
	table [y=l1, x = Q]{tab-levels.dat};
\addlegendentry{level 2}
\addplot[gray]
	table [y=l2, x = Q]{tab-levels.dat};
\end{axis}
    \end{tikzpicture}
    \qquad
    \begin{tabular}{ |c  |  c |c|} 
\hline
   $Q$ & $c_{(1,2)}^{(3)\,an}$ & $c_{(1,2)}^{(3),\,MC}$ \\
      \hline
  1  &    0.613397& 0.613208\\
  1.25  &	0.669589&0.685314\\
  1.5&	  0.719907 &0.708886\\
  1.75&	  0.766007 &0.761209 \\
  2&		0.812445 &0.811437 \\
  2.25&   0.849444&0.840513\\
  2.5&    0.88805&0.892975\\
  2.75&   0.92519&0.93614 \\
  3	&	  0.961232 &0.973289 \\
  \hline
\end{tabular}
    \captionlistentry[table]{ }
    \captionsetup{labelformat=andtable}
    \caption{Comparison of the analytic $c_{(1,2)}^{(3)}$ with the corresponding numerical coefficient, for different values of $Q$, for the isosceles geometry. In the figure we compare the coefficients computed to order $\frac{w_{12}}{w_{23}}$ (dashed) and $\left(\frac{w_{12}}{w_{23}}\right)^2$ (solid).}\label{tabfig:iso}
  \end{figure}
  

We test also the CFT predictions for triangles of different shapes. We took new Monte Carlo measurements by setting the points at:  
$(w_1,w_2,w_3) = ((k-i)r,(k+i)r,0)$.
We refer the reader to \cite{psvd13,prs16,prs19} for the details on the measurement of the three-point correlations. We compute the coefficient $c^{(3),\text{an}}_{(1,2)}=c^{(3)}_{(1,2)}\left(\frac{2}{\sqrt{k^2+1}},\tau=i\right)$ which now depends on $k$. The comparison
with the Monte Carlo measurements, taken at $Q=1$ and for different values of $k$ is shown in figure \ref{fig:3ptk}. For large $k$ we expect our $\frac{w_{12}}{w_{23}}$ expansion to converge better, however the numerical measurements get less precise for large $k$, which explains the deviation between analytical and numerical points in figure \ref{fig:3ptk}. Still, the agreement is good. 

\noindent In figure \ref{fig:R}, we plotted the ratio
\begin{equation}
R(w_1,w_2,w_3) = \frac{p_{123}(w_1,w_2,w_3)}{\sqrt{p_{12}(w_1,w_2)\,p_{23}(w_2,w_3)\,p_{13}(w_1,w_3)}}
\end{equation}
at $Q = 1$. This ratio was considered in \cite{devi11} and \cite{ziff11}. Using our expression (\ref{p3Potts}) for $p_{123}$ and the result in \cite{jps19two} for $p_{12}$,
\begin{align}\label{R}
R &=\frac{D_{(0,\frac12),(0,\frac12)}^{(0,\frac12)}\left(1+c_{1,2}^{(3)}\left|\frac{w_{23}}{N}\right|^{2\Delta_{(1,2)}}+\cdots\right)}{\displaystyle\prod_{i<j}\left(1+c_{1,2}^{(2)}\left|\frac{w_{ij}}{N}\right|^{2\Delta_{(1,2)}}+\cdots\right)^{\frac12}}\\ \nonumber
&= D_{(0,\frac12),(0,\frac12)}^{(0,\frac12)}\left[1+\left( \left|\frac{w_{23}}{r}\right|^{2\Delta_{(1,2)}}c_{1,2}^{(3)}-\frac12 c_{1,2}^{(2)}\sum_{i<j}\left|\frac{w_{ij}}{r}\right|^{2\Delta_{(1,2)}}\right)\left(\frac{r}{N}\right)^{2\Delta_{(1,2)}}+\cdots\right].
\end{align}
In particular in \cite{ziff11} the quantity $\text{ln}\left(D_{(0,\frac12),(0,\frac12)}^{(0,\frac12)}-R\right)$ was studied numerically for percolation, as a function of the log of the distance between the points. From (\ref{R}), the behaviour is,
\begin{equation}\label{DR}
\text{ln}\left(D_{(0,\frac12),(0,\frac12)}^{(0,\frac12)}-R\right)=\text{ln}\left[\frac{D_{(0,\frac12),(0,\frac12)}^{(0,\frac12)}\left( \frac12 c_{1,2}^{(2)}\sum_{i<j}\left|\frac{w_{ij}}{r}\right|^{2\Delta_{(1,2)}}-\left|\frac{w_{23}}{r}\right|^{2\Delta_{(1,2)}}c_{1,2}^{(3)}\right)}{N^{2\Delta_{(1,2)}}}\right]+2\Delta_{(1,2)}\text{ln}\, r.
\end{equation}
Then for any configuration of points the slope equals $2\Delta_{(1,2)} = 1.25$ for percolation. With the points at the vertices of an equilateral triangle, this slope was measured in  \cite{ziff11} to be $\sim 1.3$ in the regime where the distance between points is large, which is in fair agreement with our prediction. Note that for equilateral as well as isosceles triangles parametrised with $k$, the coefficient $\left( \left|\frac{w_{23}}{r}\right|^{2\Delta_{(1,2)}}c_{1,2}^{(3)}-\frac12 c_{1,2}^{(2)}\sum_{i<j}\left|\frac{w_{ij}}{r}\right|^{2\Delta_{(1,2)}}\right)$ is negative, resulting in a decrease of the ratio $R$ when $r$ approaches $N/2$.
\begin{figure}[H]\centering
\begin{subfigure}{0.4\textwidth}
\begin{tikzpicture}
\begin{axis}[
	legend cell align=center,
	xlabel={$k$},
	ylabel={$c^{(3)}_{(1,2)}\left(\frac{2}{\sqrt{k^2+1}},\tau=i\right)$},	
legend pos=south east]
	]
\addlegendentry{Monte Carlo}
\addplot[only marks,mark=o,green,mark size=1pt,error bars/.cd,y dir=both,y explicit] 
    table[x=x,y=y,y error=error] {
  x   y   error
  2	1.37		0.01
  3	1.89		0.01
  4	2.49		0.01
  5	3.13		0.01	
  6	3.81		0.01
  7	4.53		0.01
  8	5.28		0.01
  9	6.04		0.01
  10	6.80		0.02
};
\addlegendentry{Analytic}
\addplot[only marks,gray, mark= +,mark size=1pt]
coordinates{
(2, 1.33043) (3,1.83433) (4, 2.40931) (5, 3.03298) (6,3.69526) (7, 4.39004) (8,5.11308) (9, 5.8613 ) (10, 6.63228)};
\end{axis}
    \end{tikzpicture}
    \caption{Comparison of the analytic $c_{(1,2)}^{(3)}$ with the corresponding numerical coefficient, for different values of the geometric parameter $k$.}
    \label{fig:3ptk}
\end{subfigure}\hfill
\begin{subfigure}{0.4\textwidth}
\begin{tikzpicture}
\begin{semilogxaxis}[
	legend cell align=center,
	xlabel={$r/N$},
	ylabel={$\frac{p_{123}}{\sqrt{p_{12}\,p_{23}\,p_{13}}}$},	
	/pgf/number format/precision=5,
	legend pos=south west]
	]
\addplot+[green,mark=o,only marks,mark size=.6pt] 
	table [y=MC, x=r/L]{tab1-3pt-mc.dat};
\addplot[gray]
	table [y=ana, x = r/L]{tab1-3pt-ana.dat};
\legend{Monte Carlo, Analytic};
\end{semilogxaxis}
\end{tikzpicture}
\caption{Comparison of the ratio (\ref{R}) with the Monte Carlo data. At short distances the numerical point deviate significantly since this regime is not captured by the CFT description.}
    \label{fig:R}
\end{subfigure}\caption{ }
\end{figure}

\subsection{Four-point connectivity $p_{1234}$}
According to the $s-$ channel fusion for $p_{1234}(w_1,w_2,w_3,w_4)$, see \cite{js18} and \cite{prs19}, the main topological corrections are of order $O\left(\left(\frac{w_{24}}{N}\right)^{2\Delta_{(1,2)}}\right)$. We define the associated coefficient $c^{(4)}_{(1,2)}(\frac{w_{12}}{w_{24}},\frac{w_{34}}{w_{24}},\tau)$ as:
\begin{equation}\label{p1234}
p_{1234}=\frac{c_0^2}{|w_{12}w_{34}|^{4\Delta_{(0,\frac12)}}}\left[P_0\left(z\right)+ c^{(4)}_{(1,2)}\left(\frac{w_{12}}{w_{24}},\frac{w_{34}}{w_{24}},\tau\right)\left(\frac{w_{24}}{N}\right)^{2\Delta_{(1,2)}}+\text{subleading}\right],
\end{equation}
\noindent We compute the dominant contributions to the coefficient $c^{(4)}_{(1,2)}$ which, for all values of $Q$, come from the terms associated to diagrams \ref{fig:4ptconn1} and \ref{fig:4ptconn2}. Each contribution is of order $z^{\Delta_L+\Delta_R}$. 
\begin{figure}[H]\centering
 \begin{tikzpicture}[baseline=(current  bounding  box.center), very thick, scale = .5]
\draw (-3,-3) node [left] {$V_{(0,\frac12)}$}--(0,0)--(-3,3) node[left]{$V_{(0,\frac12)}$};
\draw (0,0)  -- (8,0);
\draw (2,0) node[below]{$V_{(\Delta_{(0,\frac12)},Y_L)}$};
\draw (4,0) -- (4,2.5);
\draw (4,4.3) circle(1.8cm);
\draw (4,1.5) node [right]{$V_{(1,2)}$};
\draw (6.5,0) node[below]{$V_{(\Delta_{(0,\frac12)},Y_R)}$};
\draw (11,3) node [right] {$V_{(0,\frac12)}$}--(8,0)--(11,-3) node[right]{$V_{(0,\frac12)}$};
\end{tikzpicture}
\caption{Diagrammatic representation of the leading contribution to the topological corrections of the four-point connectivity.}
\label{fig:4ptconn1}
\end{figure}

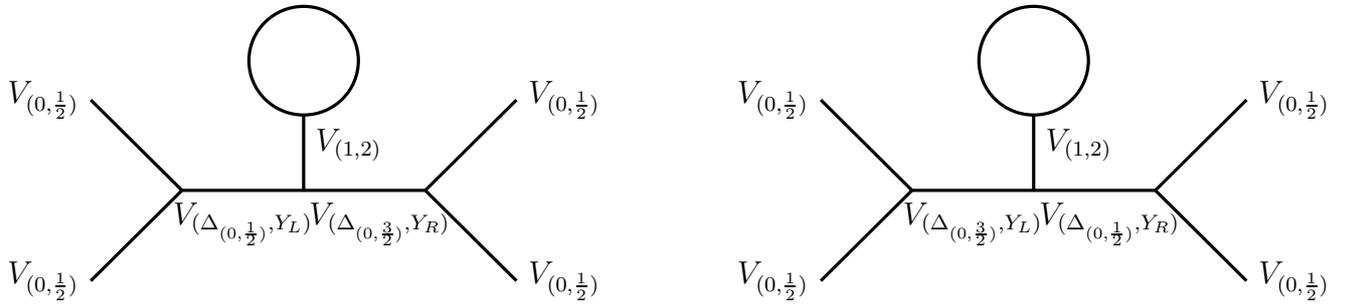
\begin{figure}[H]\centering
\begin{subfigure}{0.4\textwidth}
\begin{tikzpicture}[baseline=(current  bounding  box.center), very thick, scale = .4]
\draw (-3,-3) node [left] {$V_{(0,\frac12)}$}--(0,0)--(-3,3) node[left]{$V_{(0,\frac12)}$};
\draw (0,0)  -- (8,0);
\draw (2,0) node[below]{$V_{(\Delta_{(0,\frac12)},Y_L)}$};
\draw (4,0) -- (4,2.5);
\draw (4,4.3) circle(1.8cm);
\draw (4,1.5) node [right]{$V_{(1,2)}$};
\draw (6.5,0) node[below]{$V_{(\Delta_{(0,\frac32)},Y_R)}$};
\draw (11,3) node [right] {$V_{(0,\frac12)}$}--(8,0)--(11,-3) node[right]{$V_{(0,\frac12)}$};
\end{tikzpicture}
\end{subfigure}\hfill
\begin{subfigure}{0.4\textwidth}
 \begin{tikzpicture}[baseline=(current  bounding  box.center), very thick, scale = .4]
\draw (-3,-3) node [left] {$V_{(0,\frac12)}$}--(0,0)--(-3,3) node[left]{$V_{(0,\frac12)}$};
\draw (0,0)  -- (8,0);
\draw (2,0) node[below]{$V_{(\Delta_{(0,\frac32)},Y_L)}$};
\draw (4,0) -- (4,2.5);
\draw (4,4.3) circle(1.8cm);
\draw (4,1.5) node [right]{$V_{(1,2)}$};
\draw (6.5,0) node[below]{$V_{(\Delta_{(0,\frac12)},Y_R)}$};
\draw (11,3) node [right] {$V_{(0,\frac12)}$}--(8,0)--(11,-3) node[right]{$V_{(0,\frac12)}$};
\end{tikzpicture}
\end{subfigure}
\caption{Diagrammatic representation of the sub-leading contribution to the topological corrections of the four-point connectivity.}
\label{fig:4ptconn2}
\end{figure}
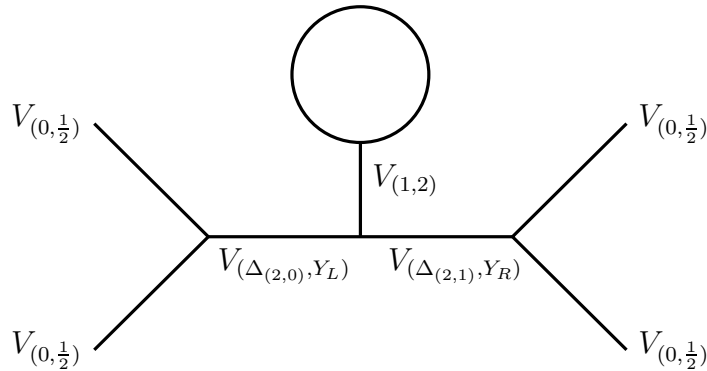
\begin{figure}[H]\centering
 \begin{tikzpicture}[baseline=(current  bounding  box.center), very thick, scale = .5]
\draw (-3,-3) node [left] {$V_{(0,\frac12)}$}--(0,0)--(-3,3) node[left]{$V_{(0,\frac12)}$};
\draw (0,0)  -- (8,0);
\draw (2,0) node[below]{$V_{(\Delta_{(2,0)},Y_L)}$};
\draw (4,0) -- (4,2.5);
\draw (4,4.3) circle(1.8cm);
\draw (4,1.5) node [right]{$V_{(1,2)}$};
\draw (6.5,0) node[below]{$V_{(\Delta_{(2,1)},Y_R)}$};
\draw (11,3) node [right] {$V_{(0,\frac12)}$}--(8,0)--(11,-3) node[right]{$V_{(0,\frac12)}$};
\end{tikzpicture}
\caption{Diagrammatic representation of the next to sub-leading contribution to the topological corrections of the four-point connectivity. This contribution is not visible numerically.}
\label{fig:4ptconn3}
\end{figure}
\begin{figure}[H]\centering
\begin{subfigure}{0.4\textwidth}
\begin{tikzpicture}
\begin{semilogxaxis}[
	legend cell align=center,
	xlabel={$r$},
	ylabel={$r^{8\Delta_{(0,\frac12)}}p_{1234}(r,z)$},	
	legend pos=north west]
	]
\addplot+[cyan,mark=o,only marks,mark size=.6pt,error bars/.cd,y dir=both,y explicit] 
	table [y=p04, x=r, y error = dp04]{table-275-4.dat};
\addplot+[red,mark=o,only marks,mark size=.6pt,error bars/.cd,y dir=both,y explicit] 
	table [y=p012, x=r, y error = dp012]{table-275-12.dat};
\addplot+[green,mark=o,only marks,mark size=.6pt,error bars/.cd,y dir=both,y explicit] 
	table [y=p032, x=r, y error = dp032]{table-275-32.dat};
\addplot[gray,forget plot]
	table [y=p04, x = r]{ana-275-4.dat};
\addplot[gray,forget plot]
	table [y=p012, x = r]{ana-275-12.dat};
\addplot[gray]
	table [y=p032, x = r]{ana-275-32.dat};
\legend{$z=0.5$,$z=0.1$,$z=0.015$,analytic};
\end{semilogxaxis}
\end{tikzpicture}\caption{}\label{fig:zs}
\end{subfigure}\hfill
\begin{subfigure}{0.4\textwidth}
\begin{tikzpicture}
\begin{semilogxaxis}[
	legend cell align=center,
	xlabel={$r$},	
	legend pos=north west]
	]
\addlegendentry{Monte Carlo}
\addplot+[blue,mark=o,only marks,mark size=.6pt,error bars/.cd,y dir=both,y explicit] 
	table [y=MC, x=r, y error = dMC]{table-275-4-levs.dat};
\addlegendentry{level 1}
\addplot[gray,dashed]
	table [y=lev1, x = r]{table-275-4-levs.dat};
\addlegendentry{level 2}
\addplot[gray]
	table [y=lev2, x = r]{table-275-4-levs.dat};
\end{semilogxaxis}
\end{tikzpicture}\caption{}\label{fig:levels}
\end{subfigure}
\caption{Numerical and analytic rescaled four-point connectivity at $Q = 2.75$, for different values of the cross-ratio $z$ (a) and for $z = 0.5$ (b) where we show the convergence of the level expansion.}
\end{figure}
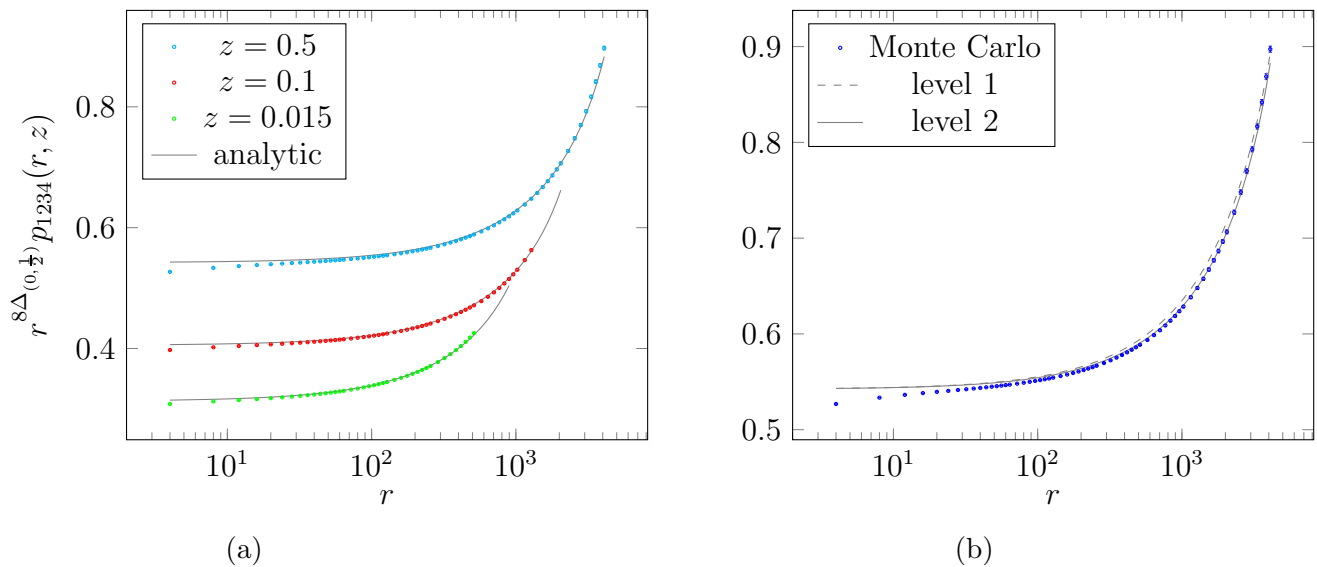
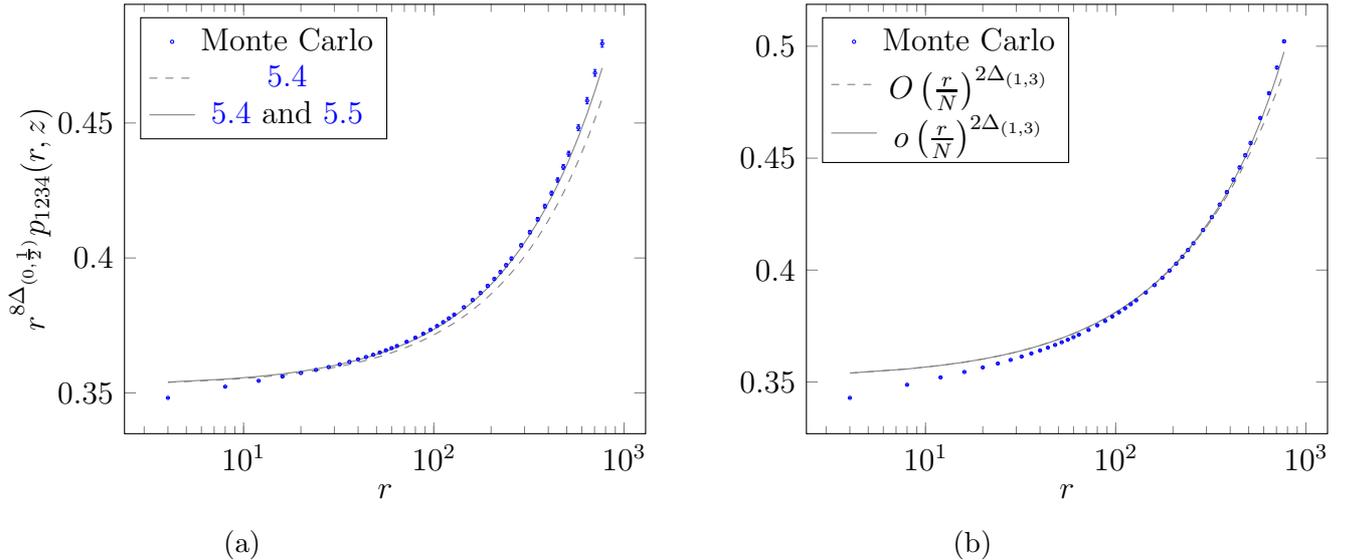
\begin{figure}[H]\centering
\begin{subfigure}{0.4\textwidth}
\begin{tikzpicture}
\begin{semilogxaxis}[
	legend cell align=center,
	xlabel={$r$},
	ylabel={$r^{8\Delta_{(0,\frac12)}}p_{1234}(r,z)$},	
	legend pos=north west]
	]
\addlegendentry{Monte Carlo}
\addplot+[blue,mark=o,only marks,mark size=.6pt,error bars/.cd,y dir=both,y explicit] 
	table [y=MC, x=r, y error = dMC]{table-275-20-diags.dat};
\addlegendentry{\ref{fig:4ptconn1}}
\addplot[gray, dashed]
	table [y=1d, x = r]{table-275-20-diags.dat};
\addlegendentry{\ref{fig:4ptconn1} and \ref{fig:4ptconn2}}
\addplot[gray]
	table [y=2d, x = r]{table-275-20-diags.dat};

\end{semilogxaxis}
\end{tikzpicture}
\caption{ }
\label{fig:diags}
\end{subfigure}\hfill
\begin{subfigure}{0.4\textwidth}
\begin{tikzpicture}
\begin{semilogxaxis}[
	legend cell align=center,
	xlabel={$r$},	
	legend pos=north west]
	]
\addplot+[blue,mark=o,only marks,mark size=.6pt,error bars/.cd,y dir=both,y explicit] 
	table [y=MC, x=r, y error = dMC]{tab-325-20.dat};
\addplot[gray, dashed]
	table [y=wt13, x = r]{tab-325-20.dat};
\addplot[gray]
	table [y=w13, x = r]{tab-325-20.dat};
	\legend{Monte Carlo, $O\left(\frac{r}{N}\right)^{2\Delta_{(1,3)}}$,$o\left(\frac{r}{N}\right)^{2\Delta_{(1,3)}}$}
\end{semilogxaxis}
\end{tikzpicture}
\caption{}
\label{fig:V13}
\end{subfigure}
\caption{Numerical and analytic rescaled four-point connectivity. In (a) we show the convergence of the diagrammatic expansion of $c_{(1,2)}^4$. In (b) we show the effect of the contribution of the sub-dominant field $V_{(1,3)}$ when $Q$ is close to $3$.}
\end{figure}

\noindent The next contribution would come from diagram \ref{fig:4ptconn3}, for which the structure constants are unknown. However this contribution would be of order\footnote{$V_{(2,1)}$ is not a diagonal field (see Appendix \ref{sec:4ptder})} $z^{\Delta_{(2,0)}+\Delta_{(2,1)}+1}$, which is very sub-dominant.
\noindent To compare our expansion of (\ref{p1234}) with the numerical simulations we take again $\tau = i$  and we fix the four points $w_1,w_2,w_3,w_4$ at the vertices of a rectangle, ie $(w_1,w_2,w_3,w_4) = (i\,r,0,\lambda\,r,(\lambda+i)r)$. The cross ratio is $$z = \frac{w_{12}w_{34}}{w_{13}w_{24}} = \frac{1}{\lambda^2+1}.$$
In figure \ref{fig:zs} we plot the function $r^{8\Delta_{(0,\frac12)}}p_{1234}(r,z)$ at $Q = 2.75$ and for different values of the cross-ratio $z$. For $z=0.5$, we show in figure \ref{fig:levels} the convergence of the level expansion (see Appendix \ref{sec:4ptder}). Taking $\lambda\geq 5$ ensures that we can truncate the expansion at level 2 and still obtain good agreement with the numerical data. In the following we will take $\lambda=5$ corresponding to $z=0.0384615$. 
\noindent In figure \ref{fig:diags}, we compute the connectivity including the contributions of the dominant \ref{fig:4ptconn1} and first sub-dominant \ref{fig:4ptconn2} diagrams in $c^{(4)}_{(1,2)}$. Note that the value $Q=2.75$ chosen for this plot is arbitrary. 
\subsubsection{$Q>2$}\label{sec:4ptg2}
When $Q>2$ the topological correction coming from the field $V_{(1,3)^D}$, while being still sub dominant, produces a visible effect. We illustrate this for $Q=3.25$ on Fig. \ref{fig:V13}, where the term $c^{(4)}_{(1,3)}\left(\frac{w_{12}}{w_{24}},\frac{w_{34}}{w_{24}},\tau\right)\left(\frac{w_{24}}{N}\right)^{2\Delta_{(1,3)}}$ is included (solid line) or not (dashed line) in the expansion of the connectivity.
\subsubsection{$Q=2$}\label{sec:Q=2}
As explained in Appendices \ref{sec:Liouville} and \ref{sec:4ptder}, some structure constants entering the computation of both the plane limit and the first topological correction diverge at $Q=2$. In particular the contribution of the $(0,\frac32)$ channel to the plane limit is divergent. As explained in Sections 2.2 and 3.3 of \cite{prs19}, for rational central charge the diverging contributions of channels with the same dimension (here $(0,\frac32)$ and $(2,0)$) cancel each other in (\ref{eq:4-ptplane1}) to give a finite limit.

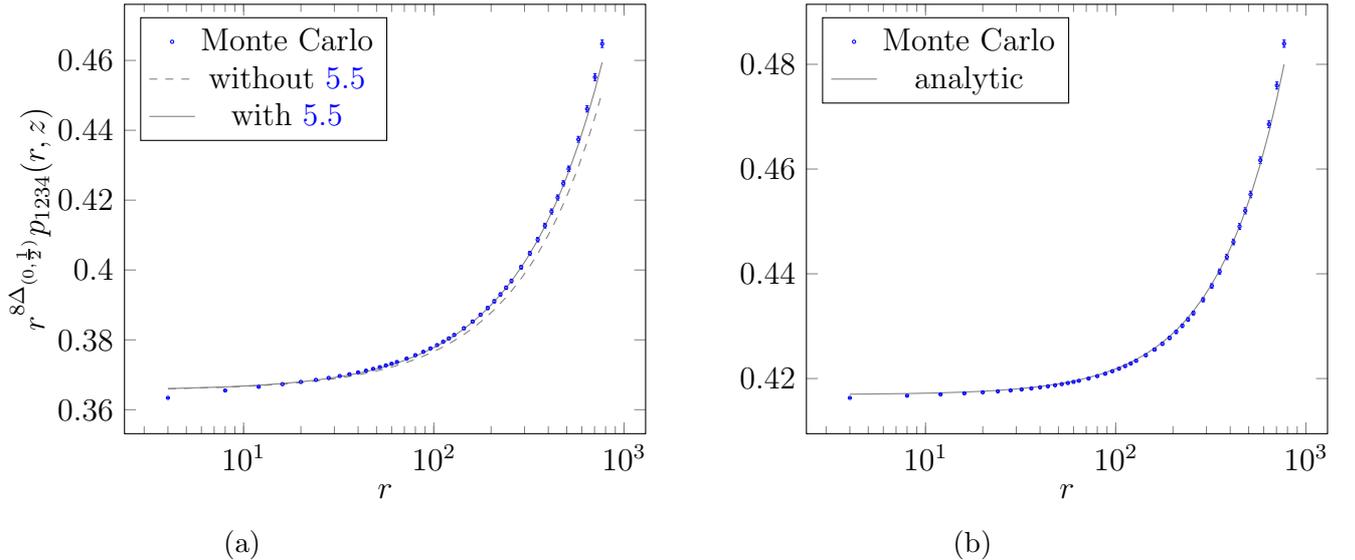
\begin{figure}[H]\centering
\begin{subfigure}{0.4\textwidth}
\begin{tikzpicture}
\begin{semilogxaxis}[
	legend cell align=center,
	xlabel={$r$},
	ylabel={$r^{8\Delta_{(0,\frac12)}}p_{1234}(r,z)$},	
	legend pos=north west]
	]
\addplot+[blue,mark=o,only marks,mark size=.6pt,error bars/.cd,y dir=both,y explicit] 
	table [y=MC, x=r, y error = dMC]{tab-2-20.dat};
\addplot[gray,dashed]
	table [y=wt32, x = r]{tab-2-20.dat};
\addplot[gray]
	table [y=w32, x = r]{tab-2-20.dat};
	\legend{Monte Carlo, without \ref{fig:4ptconn2}, with \ref{fig:4ptconn2}}
\end{semilogxaxis}
\end{tikzpicture}
\caption{ }
\label{fig:V32}
\end{subfigure}\hfill
\begin{subfigure}{0.4\textwidth}
\begin{tikzpicture}
\begin{semilogxaxis}[
	legend cell align=center,
	xlabel={$r$},	
	legend pos=north west]
	]
\addplot+[blue,mark=o,only marks,mark size=.6pt,error bars/.cd,y dir=both,y explicit] 
	table [y=MC, x=r, y error = dMC]{tab-1-20.dat};
\addplot[gray]
	table [y=ana, x = r]{tab-1-20.dat};
\legend{Monte Carlo,analytic};
\end{semilogxaxis}
\end{tikzpicture}\caption{ }\label{MCA}
\end{subfigure}\caption{Numerical and analytic rescaled four-point connectivity in the special limits $Q=2$ and $Q=1$. In (a) we show that one must include the finite $Q\to2$ limit of the contribution of the $(0,\frac{3}{2})$ channel. In (b) 
the expression involves the non-trivial $Q\to1$ limits of the structure constants, and include the contribution of $V_{(2,0)}$ in the plane four-point function.}
\end{figure}

\noindent However, as detailed in Appendix \ref{sec:4ptQ2} the contribution of the same channel $(0,\frac32)$ to the topological correction (diagram \ref{fig:4ptconn2}) has a finite limit. In figure \ref{fig:V32} we show the connectivity computed including (solid) or not (dashed) this contribution in $c_{(1,2)}^{(4)}$. The comparison with the numerical data seems to indicate that this channel must be included in the topological corrections.
\subsubsection{$Q<2$}\label{sec:Q<2}
In this section we will only consider the case $Q=1$. For other values of $Q$ the computation of the connectivity is similar to what we showed before. Note however that for $1\leq Q\leq2$, considering that $2\Delta_{(2,0)} < 2\Delta_{(0,\frac32)}$ the contribution of $(2,0)$ to the plane limit (\ref{eq:4-ptplane2}) is dominant over $(0,\frac32)$. This contribution cannot be computed using our approach since the structure constant $D_{(0,\frac12),(0,\frac12)}^{(2,0)}$ is unknown for arbitrary $Q$. Nonetheless, the contribution of this field is very small ($\sim 5.\, 10^{-5}$ at $Q=1$) and simply neglecting it gives a good agreement with the numerical data, for all $Q\in[1,4]$.
In figure \ref{MCA} we plot the connectivity at Q = 1, whose expression involves non-trivial limits of the structure constants, detailed in Appendices \ref{sec:Liouville} and \ref{sec:4ptder}. We include the contribution of $V_{(2,0)}$ in the plane four-point function (\ref{eq:4-ptplane2}) using the structure constant computed numerically in \cite{prs19}, though this contribution is very small.

\section{Conclusions}
\label{sec:conclusions}
In this paper we completed the work initiated in \cite{jps19two} where the two-point connectivity of critical Potts clusters living on a torus had been considered. We focused on the three-point and four-point connectivities $p_{123}$ and $p_{1234}$ defined in (\ref{eq:n-ptscal}). Motivated by the understanding of the CFT which describes the critical Potts clusters, the study of these higher-point connectivities is particularly interesting as it probes more fusion rules than the two-point connectivity. Moreover, contrary to the two-point connectivity, the three- and four-point connectivities cannot be written in terms of local correlation functions.
In the CFT approach, explained in Section \ref{CFTapproach}, we used the informations coming from the works \cite{js18} and \cite{prs19}. In particular we used the fusion rules (\ref{eq:OPE2pt}), (\ref{eq:ope12}) and (\ref{eq:OPE34pt}) and the $c\leq 1$ Liouville structure constants. We computed the dominant diagrams \ref{fig:channel12} and \ref{fig:channel32} for $p_{123}$ and \ref{fig:4ptconn1} - \ref{fig:4ptconn3} for $p_{1234}$. A very satisfying agreement with the corresponding Monte Carlo measurements was found. 

\noindent We showed that the leading topological corrections for $p_{123}$ and $p_{1234}$ are expected to scale with the size as $N^{-x}$, where $x=2\Delta_{(1,2)^D}$, i.e. with an exponent which is the energy scaling dimension. Note that we worked with square tori, for which the correction coming from the stress-energy tensor vanishes. For non-square tori, depending on the aspect ratio $M/N$ this latter contribution can become dominant \cite{jps19two}.  

\noindent The results presented here further support the fact that the use of $c\leq 1$ Liouville-type constants provides a very good description of Potts clusters, even when they live on a non-trivial topology. For $Q=1$ (percolation) and $Q=2$ (Ising) we showed in Sections \ref{sec:Q=2}, \ref{sec:Q<2} and in Appendices \ref{sec:3ptder} and \ref{sec:4ptder} subtle cancellations of the singularities appearing in the Liouville constants. More generally, although the CFT approach uses correlations of local fields, it remains valid for describing these geometrical objects. 

\noindent We stress the fact that, although our results are based on functions (the Liouville-type constants in this case) which are very singular in $Q$, they turn out to have a smooth dependence on $Q$ as required by  statistical physics applications, due to the aforementioned cancellations. An interesting open question is studying more systematically these fine-tuned cancellations. In particular, one can expect these cancellations to be put in relation, in the spirit of \cite{sv13}, with the logarithmic features arising from the study of the integrable structures of the lattice model. Finally, the universal results we obtained for pure percolation can be used for testing models that are conjectured to be in the same universality class, such as, for instance, the long-range percolation appearing in the study of quantum chaos \cite{Bogo07}.

\appendix 
\section{CFT definitions and notations}\label{sec:defs}
\subsection{Kinematic data}
We first recall that for a CFT on a plane $z\in (\mathbb{C}\bigcup \{\infty\})$ \cite{rib14} with $T(z)$ and $\bar{T}(\bar{z})$ the  holomorphic and anti-holomorphic component of the stress energy-tensor, the holomorphic stress-energy  modes  $L_{n}$ form the Virasoro algebra $\mathcal{V}_c$  with central charge $c$:
\begin{equation}
\label{Vir}
\left[L_{n},L_{m}\right]=(n-m)L_{n+m}+\frac{c}{12}n(n^2-1)\delta_{n,m}.
\end{equation}
The anti-holomorphic modes  $\bar{L}_{n}$ are analogously defined and form a second Virasoro algebra $\overline{\mathcal{V}}_c$, with the same central charge, that commutes with  (\ref{Vir}).
\noindent A highest-weight representation of $\mathcal{V}_c$ is labelled by the conformal dimension $\Delta$:  it contains the primary field $V_{\Delta}$, $L_{n} \ket{V_{\Delta}}=0$ for $n>0$, and its descendants, obtained by acting with the negative modes on the primary state. Given a Young diagram $Y = \{ n_1, n_2\cdots\}$, with $n_{i} \in \mathbb{N}, n_{i}\leq n_{i+1}$,  the fields 
\begin{align}
V_{\Delta}^{(Y)}=L_{-Y} \, V_{\Delta}=L_{-n_1}L_{-n_2}\cdots \;V_{\Delta}\quad (V_{\Delta}^{(\{0\})}=V_{\Delta})
\end{align} 
form a complete basis of the $\Delta$ representation.  The descendant $V^{(Y)}_{\Delta}$ has total dimension $\Delta+|Y|$, where $|Y|=\sum n_i$ is called the level of the descendant. For general $\Delta$, the number of independent descendants is therefore the number of partitions of $|Y|$. The inner product  $H_{\Delta} (Y, Y') $ between descendants is defined as:
\begin{equation}
\label{H}
H_{\Delta} \left( Y, Y' \right) = \lim_{z\to \infty} z^{2\Delta}\;
\left< V_{\Delta}(z) L_{Y}L_{-Y'} V_{\Delta}(0) \right>,
\end{equation} 
and is completely determined by the algebra (\ref{Vir}). The spectrum $\mathcal{S}$ of a CFT is formed by the representations of $\mathcal{V}_c \otimes \overline{\mathcal{V}}_c$ appearing in the theory and labelled by the holomorphic and anti-holomorphic dimensions $\Delta,\bar{\Delta}$. In order to simplify the formulas, we use the notations  $(\Delta_i)=\Delta_i,\bar{\Delta}_i$ and  $(\Delta_i,Y_i)=(\Delta_i,Y_i),(\bar{\Delta}_i,\bar{Y}_i)$. 
The $s-$ channel expansion of the four-point conformal block is also completely determined by the algebra (\ref{Vir}):
\begin{equation}\label{F4}
\mathcal{F}^{(s)}_{\Delta}(\Delta_i|z) = z^{\Delta}\left(1+\frac{(\Delta+\Delta_1-\Delta_2)(\Delta+\Delta_4-\Delta_3)}{2\Delta}z+O(z^2)\right)
\end{equation}
\subsection{Dynamic data}
The product of two fields (OPE) can be expanded in terms of the states appearing in the spectrum $\mathcal{S}$ \cite{rib14}: 
\begin{align}
\label{OPE}
V_{(\Delta_1,Y_1)}(z,\bar{z})\;V_{(\Delta_2,Y_2)}(0)& \to z^{-\Delta_1-|Y_1|-\Delta_2-|Y_2|+\Delta_3+|Y_3|}D^{(\Delta_3,Y_3)}_{(\Delta_1,Y_1),(\Delta_2,Y_2)} \;V_{(\Delta_3,Y_3)}(0),
\end{align}
where the coefficients are factorised as:
\begin{align}
\label{Dfact}
D^{(\Delta_3,Y_3)}_{(\Delta_1,Y_1),(\Delta_2,Y_2)}= D^{(\Delta_3)}_{(\Delta_1),(\Delta_2)} \;\beta^{(\Delta_3,Y)}_{(\Delta_{1},Y_1),(\Delta_2,Y_2)}\;\beta^{(\bar{\Delta}_3,\bar{Y}_3)}_{(\bar{\Delta}_1,\bar{Y}_1),(\bar{\Delta}_2,\bar{Y}_2)}.
\end{align}
One factor is fixed by the algebra (\ref{Vir}):
\begin{align}
\label{beta}
\beta^{(\Delta_3,Y_3)}_{(\Delta_{1},Y_1),(\Delta_2,Y_2)}= \sum_{\substack{Y,\\|Y|=|Y_3|}}H_{\Delta_3}^{-1}\left(Y,Y_3\right)\Gamma_{(\Delta_2,Y_2),(\Delta_1,Y_1))}^{(\Delta_3,Y)}, 
\end{align}
where the Virasoro matrix elements $\Gamma_{(\Delta_2,Y_2),(\Delta_1,Y_1)}^{(\Delta_3,Y)}$ relate three states (ie are associated to the knots of  the conformal block diagrams) and are completely determined by the commutation relation (\ref{Vir}). They can be computed using the recursion relations in \cite{kms10}.
The other factor is the (model dependent) structure constant  $D^{(\Delta_3)}_{(\Delta_1),(\Delta_2)}$. These constants can be defined as:
\begin{equation}
\label{def:D}
D_{(\Delta_1),(\Delta_2)}^{(\Delta_3)}=\left< V_{(\Delta_1)}(\infty) V_{(\Delta_2)}(1)V_{(\Delta_{2})}(0)\right>_{\text{Plane}},
\end{equation}  
where $\left<\cdots \right>_{\text{Plane}}$ is the CFT correlator on the infinite plane. The three-point functions determine the fusions between the different representations appearing in the spectrum. Note that, as also recently pointed out in \cite{dotsenko2019spins}, a more solid definition of structure constants is based on the four-point function. There can be indeed subtleties as the ones discussed in Section 5.3 of \cite{sv13}. In the case under consideration here, we can safely define the structure constants as in (\ref{def:D}). 

\subsection{One- and two- point functions on the torus}\label{subsec:12pt}
We recall here the topological expansion of the two-point function of primary or descendant (spin-less) fields :
\begin{align}\label{app2pt}
&\frac{\langle V_{(\Delta_1,Y_1)}(w_1)V_{(\Delta_2,Y_2)}(w_2)\rangle}{\left|w_{12}\right|^{-2\Delta_1-2\Delta_2}w_{12}^{-|Y_1|-|Y_2|}\bar{w}_{12}^{-|\bar{Y}_1|-|\bar{Y}_2|}}\\\nonumber
&= \sum_{(\Delta_{\text{top}},Y)}D_{(\Delta_1,Y_1),(\Delta_2,Y_2)}^{(\Delta_{\text{top}},Y)}\left(\frac{w_{12}}{N}\right)^{\Delta_{\text{top}}+|Y|}\left(\frac{\bar{w}_{12}}{N}\right)^{\bar{\Delta}_{\text{top}}+|\bar{Y}|}\langle V_{(\Delta_{\text{top}},Y)}\rangle_{(N=1)}\\\nonumber
&=\sum_{(\Delta_{\text{top}})}D_{(\Delta_1),(\Delta_2)}^{(\Delta_{\text{top}})}\left(\frac{w_{12}}{N}\right)^{\Delta_{\text{top}}}\left(\frac{\bar{w}_{12}}{N}\right)^{\bar{\Delta}_{\text{top}}} \Bigg[\beta^{(\Delta_{\text{top}})}_{(\Delta_1,Y_1),(\Delta_2,Y_2)}\beta^{(\bar{\Delta}_{\text{top}})}_{(\Delta_1,\bar{Y}_1),(\Delta_2,\bar{Y}_2)}\langle V_{(\Delta_{\text{top}})}\rangle_{(N=1)} \\\nonumber
&+ \beta^{(\Delta_{\text{top}},-1)}_{(\Delta_1,Y_1),(\Delta_2,Y_2)}\beta^{(\bar{\Delta}_{\text{top}})}_{(\Delta_1,\bar{Y}_1),(\Delta_2,\bar{Y}_2)}\langle L_{-1}V_{(\Delta_{\text{top}})}\rangle_{(N=1)} \frac{w_{12}}{N}\\\nonumber
&+\beta^{(\Delta_{\text{top}})}_{(\Delta_1,Y_1),(\Delta_2,Y_2)}\beta^{(\bar{\Delta}_{\text{top}},-1)}_{(\Delta_1,\bar{Y}_1),(\Delta_2,\bar{Y}_2)}\langle \bar{L}_{-1}V_{(\Delta_{\text{top}})}\rangle_{(N=1)} \frac{\bar{w}_{12}}{N}+O\left(\frac{w_{12}}{N}\frac{\bar{w}_{12}}{N}\right)\Bigg].
\end{align}
The torus one-point function can be expanded in the elliptic nome $q$ as \cite{fl09}:
\begin{align}\label{app1pt}
\left< V_{(\Delta,Y)}\right>_{N=1} &=\frac{(2\pi)^{\Delta+|Y|+\bar{\Delta}+|\bar{Y}|}}{Z}\sum_{(\Delta',Y')}D_{(\Delta,Y),(\Delta',Y')}^{(\Delta',Y')}q^{\Delta'-c/24+|Y'|}\bar{q}^{\bar{\Delta'}-c/24+|\bar{Y}'|}\\\nonumber
&=\frac{(2\pi)^{\Delta+|Y|+\bar{\Delta}+|\bar{Y}|}}{Z}\sum_{(\Delta')}D_{(\Delta),(\Delta')}^{(\Delta')}q^{\Delta'-c/24}\bar{q}^{\bar{\Delta'}-c/24}\Bigg|1+\beta_{(\Delta,Y),(\Delta',-1)}^{(\Delta',-1)}q\\\nonumber
&+\left(\beta_{(\Delta,Y),(\Delta',\{-1,-1\})}^{(\Delta',\{-1,-1\})}+\beta_{(\Delta,Y),(\Delta',-2)}^{(\Delta',-2)}\right)q^2+\cdots\Bigg|^2.
\end{align}
%
\section{$c\leq1$ Liouville structure constants}\label{sec:Liouville}
The Liouville structure constants are the unique solutions of certain bootstrap equations for central charge $c\leq1$ \cite{RiSa14}. The structure constant of fields with dimensions $\Delta_1,\Delta_2,\Delta_3$ is given by \cite{zam05} \cite{sch05} \cite{kp05a}:
\begin{equation}\label{defC}
C_{(\Delta_1),(\Delta_2)}^{(\Delta_3)} = -A(\beta)\frac{\Upsilon_\beta(\alpha_1+\alpha_2+\alpha_3+2\beta-1/\beta)\prod_{\sigma\in\mathcal{S}_3}\Upsilon_\beta(\alpha_{\sigma(1)}+\alpha_{\sigma(2)}-\alpha_{\sigma(3)}+\beta)}{\sqrt{\displaystyle\prod_{j=1}^3\Upsilon_\beta(2\alpha_j+\beta)\Upsilon_\beta(2\alpha_j+2\beta-1/\beta)}}
\end{equation}
where the charges $\alpha$ are related to the dimensions $\Delta$ by $\Delta = i\alpha(\frac{c-1}{24}-i\alpha)$, and,
\begin{equation}
A(\beta) = \frac{\beta^{\beta^{-2}-\beta^2-1}}{\Upsilon_\beta(\beta)}\sqrt{\gamma(\beta^2)\gamma(\beta^{-2}-1)},\quad \gamma(x) = \frac{\Gamma(x)}{\Gamma(1-x)}.
\end{equation}
The special function $\Upsilon_\beta$ obeys the shift equations,
\begin{subequations}
\begin{align}
&\Upsilon_\beta(x+\beta) = \Upsilon_\beta(x)\beta^{1-2\beta x}\gamma(\beta x)\\
&\Upsilon_\beta(x+\frac{1}{\beta}) = \Upsilon_\beta(x)\beta^{2\frac{x}{\beta}-1}\gamma(\frac{x}{\beta})\\
&\Upsilon_\beta(\beta+\frac{1}{\beta}-x) = \Upsilon_\beta(x).
\end{align}
\end{subequations}
$\Upsilon_\beta$ has simple zeroes for $x\in\left(-\beta \mathds{N}-\frac{1}{\beta}\mathds{N}\right)\cup\left(\beta\mathds{N}_*+\frac{1}{\beta}\mathds{N}_*\right)$. When one of the fields is degenerate, (\ref{defC}) can be written in terms of Gamma functions.
\noindent Some constants entering the computation of the connectivities of the $Q-$ Potts model become singular for special values of $Q$:
\begin{subequations}\label{sc}
\begin{align}
C_{(0,\frac12),(0,\frac12)}^{(1,2)^D} &= -4\beta^4\frac{\Gamma(1+\frac12\beta^{-2})}{\Gamma(-\frac12\beta^{-2})}\sqrt{\frac{2^{3-4\beta^{-2}}\Gamma(\frac32-\beta^{-2})}{\Gamma(-\frac12+\beta^{-2})}}\label{scsse}\\
&=\frac49 \frac{\Gamma(\frac74)}{\Gamma(\frac14)}\sqrt{\frac{\pi\sqrt{3}}{Q-1}},\quad Q\to 1\nonumber\\
C_{(0,\frac12),(0,\frac32)}^{(1,2)^D}&=\beta^2\left[-\frac{\Gamma(2-2\beta^{-2})\Gamma(-\frac{1}{2}\beta^{-2})\Gamma(\beta^{-2})\Gamma(\frac32 \beta^{-2})}{\Gamma(1-\frac32\beta^{-2})\Gamma(1+\frac12\beta^{-2})\Gamma(-1+2\beta^{-2})\Gamma(-\beta^{-2})}\right]^{1/2} \label{scsze}\\
&=-\frac{1}{3}\sqrt{\frac{\pi\sqrt{3}}{Q-1}\frac{\Gamma(-\frac34)\Gamma(\frac94)}{\Gamma(-\frac54)\Gamma(\frac74)}},\quad Q\to 1\nonumber \\
&\propto 1/\sqrt{\Gamma(1-\frac32\beta^{-2})}\sim \sqrt{Q-2},\quad Q\to2\nonumber\\
C_{(0,\frac12),(0,\frac12)}^{(0,\frac32)} & \propto \sqrt{\Gamma(1-\frac32\beta^{-2})}\sim\frac{1}{\sqrt{Q-2}}.\label{scssz}
\end{align}
\end{subequations}

\section{Derivation of the three-point corrections}\label{sec:3ptder}
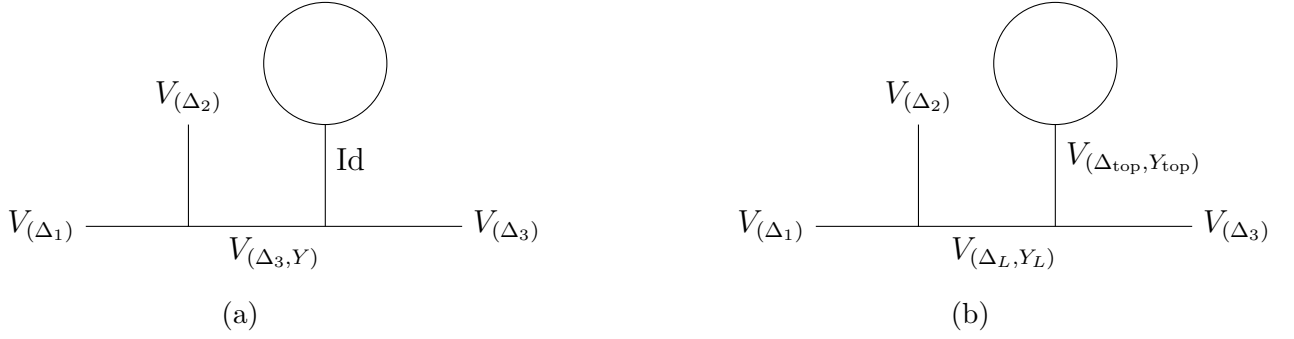
\begin{figure}[H]\centering
\begin{subfigure}{0.4\textwidth}
\begin{tikzpicture}[scale = .45]
\draw (-3,0) node [left] {$V_{(\Delta_1)}$}--(0,0)--(0,3) node[above]{$V_{(\Delta_2)}$};
\draw (0,0)  -- (8,0);
\draw (2.5,0) node[below]{$V_{(\Delta_3,Y)}$};
\draw (4,0) -- (4,3);
\draw (4,4.8) circle(1.8cm);
\draw (4,2) node [right] {$\text{Id}$};
\draw (8,0) node [right] {$V_{(\Delta_3)}$};
\end{tikzpicture}
\caption{ }
\label{appfig:3ptid}
\end{subfigure}\hfill
\begin{subfigure}{0.4\textwidth}
\begin{tikzpicture}[scale = .45]
\draw (-3,0) node [left] {$V_{(\Delta_1)}$}--(0,0)--(0,3) node[above]{$V_{(\Delta_2)}$};
\draw (0,0)  -- (8,0);
\draw (2.5,0) node[below]{$ V_{(\Delta_L,Y_L)}$};
\draw (4,0) -- (4,3);
\draw (4,4.8) circle(1.8cm);
\draw (4,2) node [right] {$V_{(\Delta_{\text{top}},Y_{\text{top}})}$};
\draw (8,0) node [right] {$V_{(\Delta_3)}$};
\end{tikzpicture}
\caption{ }
\label{appfig:3ptt}
\end{subfigure}
\caption{Diagrammatic representation of the plane limit (a) and the topological corrections (b) of the torus three-point function.}
\end{figure}

We write the s-channel expansion of the three-point function $\langle V_{(\Delta_1)}(w_1)V_{(\Delta_2)}(w_2)V_{(\Delta_3)}(w_3)\rangle$ of spin-less fields\footnote{for simplicity of notation we derive the result for spin-less fields; it is straightforward to extend it to fields with spin.} by inserting the OPE $V_{(\Delta_1)}(w_1)V_{(\Delta_2)}(w_2)$:
 \begin{multline}\label{appeq:3pt}
\frac{\langle V_{(\Delta_1)}(w_1)V_{(\Delta_2)}(w_2)V_{(\Delta_3)}(w_3)\rangle}{\left|w_{12}\right|^{-2\Delta_1-2\Delta_2}}=\sum_{\substack{(\Delta_L)\in \mathcal{S}\\(Y_L)}}D_{(\Delta_1),(\Delta_2)}^{(\Delta_L,Y_L)}w_{12}^{\Delta_L+|Y_L|}\bar{w}_{12}^{\bar{\Delta}_L+|\bar{Y}_L|} \left< V_{(\Delta_L,Y_L)}(w_2)V_{(\Delta_3)}(w_3)\right>.
\end{multline}
The plane limit is given by the term $\Delta_L = \Delta_3$ corresponding to diagram \ref{appfig:3ptid} while the topological corrections are associated to diagrams \ref{appfig:3ptt}.
Accordingly we write (\ref{appeq:3pt}) as,
 \begin{multline}
\frac{\langle V_{(\Delta_1)}(w_1)V_{(\Delta_2)}(w_2)V_{(\Delta_3)}(w_3)\rangle}{\left|w_{12}\right|^{-2\Delta_1-2\Delta_2}\left|w_{23}\right|^{2\Delta_3}}=\left|\frac{w_{12}}{w_{23}}\right|^{2\Delta_3}\sum_{(Y)}D_{(\Delta_1),(\Delta_2)}^{(\Delta_3,Y)}D_{(\Delta_3,Y),(\Delta_3)}^{\text{Id}}\left(\frac{w_{12}}{w_{23}}\right)^{|Y|}\left(\frac{\bar{w}_{12}}{\bar{w}_{23}}\right)^{|\bar{Y}|}\\
+\sum_{\substack{(\Delta_L,Y_L)\\(\Delta_{\text{top}},Y_{\text{top}})}} D_{(\Delta_1),(\Delta_2)}^{(\Delta_L,Y_L)}D_{(\Delta_L,Y_L),(\Delta_3)}^{(\Delta_{\text{top}},Y_{\text{top}})}\left(\frac{w_{12}}{w_{23}}\right)^{\Delta_L+|Y_L|}\left(\frac{\bar{w}_{12}}{\bar{w}_{23}}\right)^{\bar{\Delta}_L+|\bar{Y}_L|}\\
\times \left(\frac{w_{23}}{N}\right)^{\Delta_{\text{top}}+|Y_{\text{top}}|}\left(\frac{\bar{w}_{23}}{N}\right)^{\bar{\Delta}_{\text{top}}+|\bar{Y}_{\text{top}}|}\left< V_{(\Delta_{\text{top}},Y_{\text{top}})}\right>_{N=1}.
\end{multline}
Let us detail how to recover the plane limit from the first sum. We compute,
\begin{multline}\label{derpl}
\sum_{Y}\beta_{\Delta_1,\Delta_2}^{(\Delta_3,Y)}\beta_{\Delta_1,\Delta_2}^{(\Delta_3,\bar{Y})}\beta_{(\Delta_3,Y),\Delta_3}^{\text{Id}}\beta_{(\Delta_3,\bar{Y}),\Delta_3}^{\text{Id}}\left(\frac{w_{12}}{w_{23}}\right)^{|Y|}\left(\frac{\bar{w}_{12}}{\bar{w}_{23}}\right)^{|\bar{Y}|}\\
=\left|1+\beta_{\Delta_1,\Delta_2}^{(\Delta_3,-1)}\beta_{(\Delta_3,-1),\Delta_3}^{\text{Id}}\frac{w_{12}}{w_{23}}+\left(\beta_{\Delta_1,\Delta_2}^{(\Delta_3,\{-1,-1\})}\beta_{(\Delta_3,\{-1,-1\}),\Delta_3}^{\text{Id}}+\beta_{\Delta_1,\Delta_2}^{(\Delta_3,-2)}\beta_{(\Delta_3,-2),\Delta_3}^{\text{Id}}\right)\frac{w_{12}^2}{w_{23}^2}+\cdots\right|^2.
\end{multline}
Computing the coefficients using (\ref{beta}) and the relations in \cite{kms10} we find
\begin{align}
&\left|1-(\Delta_1-\Delta_2+\Delta_3)\frac{w_{12}}{w_{23}} +\frac{1}{2}(\Delta_1-\Delta_2+\Delta_3)(1+\Delta_1-\Delta_2+\Delta_3) \frac{w_{12}^2}{w_{23}^2}+\cdots\right|^2\\\nonumber
&=\left|1 + \frac{w_{12}}{w_{23}}\right|^{-2\Delta_1 + 2\Delta_2 - 2\Delta_3} = \left|\frac{w_{23}}{w_{13}}\right|^{2\Delta_1-2\Delta_2+2\Delta_3}
\end{align}
and therefore
\begin{multline}
\langle V_{(\Delta_1)}(w_1)V_{(\Delta_2)}(w_2)V_{(\Delta_3)}(w_3)\rangle=\frac{1}{\left|w_{12}\right|^{2\Delta_1+2\Delta_2}\left|w_{23}\right|^{2\Delta_3}}\Bigg[\frac{D_{(\Delta_1),(\Delta_2)}^{(\Delta_3)}}{\left|w_{12}\right|^{-2\Delta_3}\left|w_{23}\right|^{-2\Delta_1+2\Delta_2}\left|w_{13}\right|^{2\Delta_1-2\Delta_2+2\Delta_3}}\\
+f_\tau^{(3)}\left(\frac{w_{12}}{w_{23}},\frac{w_{23}}{N}\right)\Bigg]
\end{multline}
with the function $f_\tau^{(3)}$ defined in (\ref{def:f3}):
\begin{multline}
f_\tau^{(3)}\left(\frac{w_{12}}{w_{23}},\frac{w_{23}}{N}\right)=\sum_{(\Delta_{\text{top}},Y_{\text{top}})}c^{(3)}_{(\Delta_{\text{top}},Y_{\text{top}})}\left(\frac{w_{12}}{w_{23}}\right)\left(\frac{w_{23}}{N}\right)^{\Delta_{\text{top}}+|Y_{\text{top}}|}\left(\frac{\bar{w}_{23}}{N}\right)^{\bar{\Delta}_{\text{top}}+|\bar{Y}_{\text{top}}|}
\end{multline}
and $c^{(3)}_{(\Delta_{\text{top}},Y_{\text{top}})}\left(\frac{w_{12}}{w_{23}}\right)$ given in (\ref{defc3}).
\noindent Specialising to the $Q -$ Potts model, we took $\Delta_1=\Delta_2=\Delta_3=\Delta_{(0,\frac12)}$ and we computed the most dominant $1/N$ correction to the plane three-point function, corresponding to $\Delta_{\text{top}}=\Delta_{\text{min}} = \Delta_{(1,2)}$. As explained in Section \ref{sec:3pt}, the fusion (\ref{eq:ope12}) of $V_{(1,2)}$ imposes that either $\Delta_L = \Delta_{(0,\frac12)}$ or $\Delta_L = \Delta_{(0,\frac32)}$ corresponding to diagrams \ref{fig:channel12} and \ref{fig:channel32}. The level expansion of $c^{(3)}_{(1,2)}$ is similar to (\ref{derpl}) and was also carried out to level 2 ie $|Y_L| = 2$ in (\ref{defc3}), which showed sufficient precision for comparison with the numerical results:
\begin{align}\label{c312}
c^{(3)}_{(1,2)}\left(\frac{w_{12}}{w_{23}}\right) &= \langle V_{(1,2)}\rangle_{(N=1)}\\\nonumber
\times\Bigg\{&D_{(0,\frac12),(0,\frac12)}^{(0,\frac12)}D_{(0,\frac12),(0,\frac12)}^{(1,2)}\left|\frac{w_{12}}{w_{23}}\right|^{2\Delta_{(0,\frac12)}}\left|1+\beta_{(0,\frac12),(0,\frac12)}^{(0,\frac12,-1)}\beta_{(0,\frac12,-1),(0,\frac12)}^{(1,2)}\frac{w_{12}}{w_{23}}+O\left(\frac{w_{12}^2}{w_{23}^2}\right)\right|^2\\\nonumber
+&D_{(0,\frac12),(0,\frac12)}^{(0,\frac32)}D_{(0,\frac32),(0,\frac12)}^{(1,2)}\left|\frac{w_{12}}{w_{23}}\right|^{2\Delta_{(0,\frac32)}}\left|1+\beta_{(0,\frac12),(0,\frac12)}^{(0,\frac32,-1)}\beta_{(0,\frac32,-1),(0,\frac12)}^{(1,2)}\frac{w_{12}}{w_{23}}+O\left(\frac{w_{12}^2}{w_{23}^2}\right)\right|^2\Bigg\}
\end{align}
\subsection{Special cases: $Q=1$ and $Q=2$}
\begin{itemize}
\item When $Q=1$, the $\frac{1}{\sqrt{Q-1}}$ singularities in $D_{(0,\frac12),(0,\frac12)}^{(1,2)^D}$ and in $D_{(0,\frac12),(0,\frac32)}^{(1,2)^D}$ (resp. diagrams \ref{fig:channel12} and \ref{fig:channel32}) are cancelled in (\ref{c312}) by the factor $\sqrt{Q-1}$ in the energy one-point function
\begin{equation}
\langle V_{(1,2)^D} \rangle_{N=1} = \frac{Q-1}{Z_Q}D_{(0,\frac12),(0,\frac12)}^{(1,2)^D}\left|q\right|^{2\left(\Delta_{(0,\frac12)}-\frac{c}{24}\right)}\left|1+O(q)\right|^2,
\end{equation}
yielding a finite, non-zero limit for $c_{(1,2)}^{(3)}$.
\item When $Q=2$, the zero and the pole in (\ref{scsze}) and (\ref{scssz}) cancel in the product $C_{(0,\frac12),(0,\frac12)}^{(0,\frac32)}\,C_{(0,\frac32),(0,\frac12)}^{(1,2)^D}$, giving a finite contribution of the $(0,\frac32)$ channel \ref{fig:channel32} to $c_{(1,2)}^{(3)}$.
\end{itemize}

\section{Derivation of the four-point corrections}\label{sec:4ptder}
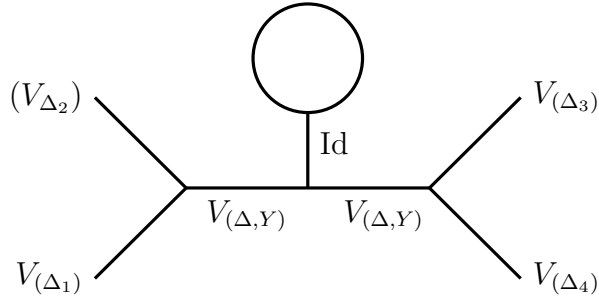
\begin{figure}[H]\centering
 \centering
 \begin{tikzpicture}[baseline=(current  bounding  box.center), very thick, scale = .4]
\draw (-3,-3) node [left] {$V_{(\Delta_1)}$}--(0,0)--(-3,3) node[left]{$(V_{\Delta_2})$};
\draw (0,0)  -- (8,0);
\draw (2,0) node[below]{$V_{(\Delta,Y)}$};
\draw (4,0) -- (4,2.5);
\draw (4,4.3) circle(1.8cm);
\draw (4,1.5) node [right]{$\text{Id}$};
\draw (6.5,0) node[below]{$V_{(\Delta,Y)}$};
\draw (11,3) node [right] {$V_{(\Delta_3)}$}--(8,0)--(11,-3) node[right]{$V_{(\Delta_4)}$};
\end{tikzpicture}
\caption{Diagrammatic representation of the plane limit of the torus four-point function.}
\label{appfig:4ptP}
\end{figure}

We write the s-channel expansion of the four-point function $\langle V_{(\Delta_1)}(w_1)V_{(\Delta_2)}(w_2)V_{(\Delta_3)}(w_3)V_{(\Delta_4)}(w_4)\rangle$ of four (primary, spin-less) fields by inserting the OPEs of $V_{(\Delta_1)}(w_1)V_{(\Delta_2)}(w_2)$ and $V_{(\Delta_3)}(w_3)V_{(\Delta_4)}(w_4)$:
 \begin{multline}
\frac{V_{(\Delta_1)}(w_1)V_{(\Delta_2)}(w_2)V_{(\Delta_3)}(w_3)V_{(\Delta_4)}(w_4)}{\left|w_{12}\right|^{-2\Delta_1-2\Delta_2}\left|w_{34}\right|^{-2\Delta_3-2\Delta_4}}=\sum_{(\Delta_L,Y_L)} D_{(\Delta_1),(\Delta_2)}^{(\Delta_L,Y_L)}w_{12}^{\Delta_L+|Y_L|}\bar{w}_{12}^{\bar{\Delta}_L+|\bar{Y}_L|}V_{(\Delta_L,Y_L)}(w_2)\\\times\sum_{(\Delta_R,Y_R)} D_{(\Delta_3),(\Delta_4)}^{(\Delta_R,Y_R)}w_{34}^{\Delta_R+|Y_R|}\bar{w}_{34}^{\bar{\Delta}_R+|\bar{Y}_R|}V_{(\Delta_R,Y_R)}(w_4).
\end{multline}
Inserting the expansion (\ref{app2pt}) of $\langle V_{(\Delta_L,Y_L)}(w_2)V_{(\Delta_R,Y_R)}(w_4)\rangle$,
 \begin{multline}
\frac{\langle V_{(\Delta_1)}(w_1)V_{(\Delta_2)}(w_2)V_{(\Delta_3)}(w_3)V_{(\Delta_4)}(w_4)\rangle }{\left|w_{12}\right|^{-2\Delta_1-2\Delta_2}\left|w_{34}\right|^{-2\Delta_3-2\Delta_4}}\\= \sum_{\substack{(\Delta_L,Y_L)\\(\Delta_R,Y_R)}}D_{(\Delta_1),(\Delta_2)}^{(\Delta_L,Y_L)}D_{(\Delta_3),(\Delta_4)}^{(\Delta_R,Y_R)}D_{(\Delta_L,Y_L),(\Delta_R,Y_R)}^{(\Delta,Y)}\left(\frac{w_{12}}{w_{24}}\right)^{\Delta_L+|Y_L|}\left(\frac{\bar{w}_{12}}{\bar{w}_{24}}\right)^{\bar{\Delta}_L+|\bar{Y}_L|}\\\times\left(\frac{w_{34}}{w_{24}}\right)^{\Delta_R+|Y_R|}\left(\frac{\bar{w}_{34}}{\bar{w}_{24}}\right)^{\bar{\Delta}_R+|\bar{Y}_R|}\sum_{(\Delta_{\text{top}},Y_{\text{top}})}\left(\frac{w_{24}}{N}\right)^{\Delta_{\text{top}}+|Y_{\text{top}}|}\left(\frac{\bar{w}_{24}}{N}\right)^{\bar{\Delta}_{\text{top}}+|\bar{Y}_{\text{top}}|}\langle V_{(\Delta_{\text{top}},Y_{\text{top}})}\rangle.
\end{multline}
The plane limit $P_0$ is given by the terms with $\Delta_{\text{top}} = 0$ and $\Delta_L = \Delta_R$, corresponding to diagrams \ref{appfig:4ptP}.
\noindent and can be written as a function of the cross-ratio $z$:
\begin{equation}\label{app:P0}
P_0(z) = \sum_{(\Delta)} D_{(\Delta_1),(\Delta_2)}^{(\Delta)}D_{(\Delta_3),(\Delta_4)}^{(\Delta)} \left|\mathcal{F}^{(s)}_{(\Delta)}(\Delta_i|z)\right|^2
\end{equation}
where $\mathcal{F}^{(s)}_{(\Delta)}(\Delta_i|z)$ is the $s-$ channel four-point conformal block (\ref{F4}). Then,
\begin{equation}
\langle V_{(\Delta_1)}(w_1)V_{(\Delta_2)}(w_2)V_{(\Delta_3)}(w_3)V_{(\Delta_4)}(w_4)\rangle=\frac{1}{\left|w_{12}\right|^{2\Delta_1+2\Delta_2}\left|w_{34}\right|^{2\Delta_3+2\Delta_4}}\left[P_0(z)+f^{(4)}_{\tau}\left(\frac{w_{12}}{w_{24}},\frac{w_{34}}{w_{24}},\frac{w_{24}}{N}\right)\right]
\end{equation}
with
\begin{equation}
f^{(4)}_{\tau}\left(\frac{w_{12}}{w_{24}},\frac{w_{34}}{w_{24}},\frac{w_{24}}{N}\right)
=\sum_{(\Delta_{\text{top}},Y_{\text{top}})}c^{(4)}_{(\Delta_{\text{top}},Y_{\text{top}})}\left(\frac{w_{12}}{w_{24}},\frac{w_{34}}{w_{24}},\tau\right) \left(\frac{w_{24}}{N}\right)^{\Delta_{\text{top}}+|Y_{\text{top}}|}\left(\frac{\bar{w}_{24}}{N}\right)^{\bar{\Delta}_{\text{top}}+|\bar{Y}_{\text{top}}|},
\end{equation}
where $c^{(4)}_{(\Delta_{\text{top}},Y_{\text{top}})}\left(\frac{w_{12}}{w_{24}},\frac{w_{34}}{w_{24}},\tau\right)$ is given by (\ref{defc4}).
\noindent The contribution of each diagram of the type in figure \ref{fig:4pt} to $c^{(4)}$ is of order $z^{\frac12(\Delta_L+\bar{\Delta}_L+\Delta_R+\bar{\Delta}_R)} = z^{\Delta_L+\Delta_r-\frac12(s_L+s_R)}$. The non-diagonal fields in the spectrum of the $Q-$ Potts model have spins $S_{(r,s)} = -rs$: those with non-zero spin give therefore very sub-dominant contributions to the four-point connectivity.
\subsection{Special cases: $Q=1$ and $Q=2$}\label{sec:4ptQ2}
\begin{itemize}
\item When $Q=1$, the $\frac{1}{\sqrt{Q-1}}$ singularities in $D_{(0,\frac12),(0,\frac12)}^{(1,2)^D}$ and in $D_{(0,\frac12),(0,\frac32)}^{(1,2)^D}$ (resp. diagrams \ref{fig:4ptconn1} and \ref{fig:4ptconn2})  are cancelled by the factor $\sqrt{Q-1}$ in the energy one-point function, exactly as in the three-point case.
\item When $Q=2$, again as in the three-point case, the zero and the pole in (\ref{scsze}) and (\ref{scssz}) coming from the contribution of diagram \ref{fig:4ptconn2} cancel in the product $C_{(0,\frac12),(0,\frac12)}^{(0,\frac32)}\,C_{(0,\frac32),(0,\frac12)}^{(1,2)^D}$. The contribution of the $(0,\frac32)$ channel to $c_{(1,2)}^{(4)}$ is therefore finite, contrary to the contribution of the same channel to the plane four-point function. In that latter case, the divergences of the different channels with the same dimension cancel each other.
\end{itemize}

\acknowledgments{{We thank Sylvain Ribault and Vladimir Dotsenko for discussion, as well as Yacine Ikhlef, Benoit Estienne and Jesper Jacobsen. R.S. thanks the International Institute of Physics of Natal for its hospitality during the period where this work was completed.}}

\bibliographystyle{morder6}
\bibliography{newletterbib}
\end{document}